\documentclass[12pt]{article}
\usepackage{jheppub}

\usepackage{subcaption}
\usepackage{xcolor}
\usepackage{amsmath}
\usepackage{amssymb}
\usepackage{amscd}
\usepackage{dsfont}
\usepackage{enumerate}
\usepackage{enumitem}
\usepackage{amsfonts}
\usepackage{epsfig}
\usepackage{mathtools}
\usepackage{yfonts}
\usepackage{bbold}
\usepackage{breqn}
\usepackage[utf8]{inputenc}
\usepackage[english]{babel}
\usepackage[subpreambles=true]{standalone}
\usepackage{import}
\usepackage{caption}
\usepackage{multirow}

\def\eqa{\begin{eqnarray}}
\def\eqae{\end{eqnarray}}
\def\eq{\begin{equation}}
\def\eqe{\end{equation}}
\def\be{\begin{equation}}
\def\ee{\end{equation}}
\def\bea{\begin{eqnarray}}
\def\eea{\end{eqnarray}}
\def\ba{\begin{array}}
\def\ea{\end{array}}

\title{Holographic Renyi entropy  of 2d CFT in KdV generalized ensemble}
\author[a,d]{Liangyu Chen,} 
\author[b]{Anatoly Dymarsky,} 
\author[a,c]{Jia Tian,}
\author[a]{and Huajia Wang}
\affiliation[a]{Kavli Institute for Theoretical Sciences, University of Chinese Academy of Sciences,
Beijing 100190, China}
\affiliation[b]{Department of Physics and Astronomy, University of Kentucky, Lexington, 40506}
\affiliation[c]{State Key Laboratory of Quantum Optics and Quantum Optics Devices, Institute of Theoretical Physics, Shanxi University, Taiyuan 030006, P.~R.~China}
\affiliation[d]{Yau Mathematical Sciences Center, Tsinghua University, Beijing 100084, China}
\emailAdd{liangyu-chen@mail.tsinghua.edu.cn}
\emailAdd{a.dymarsky@uky.edu}
\emailAdd{wukongjiaozi@ucas.ac.cn}
\emailAdd{wanghuajia@ucas.ac.cn}
\abstract 
{The eigenstate thermalization hypothesis (ETH) in chaotic two dimensional CFTs is subtle due to infinitely many conserved KdV charges. Previous works have demonstrated that  primary CFT  eigenstates have flat  entanglement spectrum, which is very different from the   microcanonical ensemble. This result is an apparent contradiction to conventional ETH, which does not take KdV charges into account.  In a companion paper \cite{KdVETHgeneral}, we resolve this discrepancy by studying the subsystem entropy of a chaotic CFT in KdV-generalized Gibbs and microcanonical ensembles. In this paper, we carry out parallel computations in the context of AdS/CFT. We focus on the high density limit, which is equivalent to  thermodynamic limit 
in conformal theories.  In this limit holographic Renyi entropy can be computed using the so-called gluing construction. We explicitly study the 
KdV-generalized microcanonical ensemble with the densities of the first two KdV charges $\langle \mathcal{Q}_1\rangle = q_1,\langle \mathcal{Q}_3\rangle = q_3$ fixed and obeying $q_3-q_1^2 \ll q_1^2$. In this regime we found that the refined Renyi entropy $\tilde{S}_n$ is $n$-independent for $n>n_{cut}$, where $n_{cut}$ depends on $q_1,q_3$. By taking the primary state limit $q_3\to q_1^2$, we recover flat entanglement spectrum characteristic of fixed-area states, in agreement  with the primary state behavior. This provides a consistency check of the KdV-generalized ETH in 2d CFTs.}

\begin{document}
\maketitle
\section{Introduction}\label{sec:intro}
Understanding the phenomena of pure state thermalization has been a crucial endeavor that, apart from its own interest and importance, plays important roles across many subjects ranging from quantum information to black hole physics -- particularly the black hole information paradox \cite{Hawking:1975IP,Hawking:1976IP,Schack:1996,Zurek:1994,Shenker:2014,Sachdev:1993,Kitaev:2015}.  A conjecture about the underlying mechanism is the notion of the eigenstate thermalization hypothesis (ETH), which proposes that high energy eigenstate whose energy density remains finite in the thermodynamic limit behave like thermal states upon evaluating the expectation values of observables \cite{Srednicki:1994,Deutsch:1991,Rigol:2008,Alessio:2016}. More precisely, in terms of the matrix elements in the eigenstate basis, ETH proposes that: 
\be\label{eq:ETH}
\langle E_a |\mathcal{O}_{obs}|E_b\rangle = f(E) \delta_{ab}+ e^{-S\left(\bar{E}\right)/2}R_{ab},\;\;\bar{E}=\frac{E_a+E_b}{2} 
\ee
where $f(E)$ is a continuous function of $E$ encoding the thermal expectation value, while the second exponentially suppressed term exhibits random matrix behavior. 

In practice, it is difficult to describe explicitly what constitutes the good observables $\mathcal{O}_{obs}$ that satisfy (\ref{eq:ETH}). For this reason, an alternative characterization of the ETH has been put forward in terms of the reduced density matrices (RDM) of the subsystems \cite{Dymarsky:2018}. In these versions, ETH proposes the proximity between the reduced density matrices $\rho^A_a = \text{Tr}_{\bar{A}}|E_a\rangle \langle E_a|$ of a high energy eigenstate $|E_a\rangle$ to those of the micro-canonical  ensembles: 
\be
\rho^A_a \approx \rho^{\text{micro}}_A 
\ee
More precisely, the notion of proximity is stated in terms of the trace distance measures between matrices:
\be
||\rho^A_a - \rho^{\text{micro}}_A || \sim \mathcal{O}\left(\Delta E/E\right),\;\;\;|| O|| = \frac{1}{2}\text{Tr}\sqrt{O O^{\dagger}}
\ee
where $\Delta E$ is the width of the energy window in defining the microcanonical ensemble. Additional support based on numerical evidence was performed in \cite{Grover:2018}.  

The notion of ETH is associated with the thermodynamic limit, i.e. a large number of degrees of freedom. While the standard thermodynamic limit is reached by taking the total system size $L$ to be large, in the context of conformal field theories (CFTs) one can explicitly define an ``internal" thermodynamic limit in which the central charge $c$ becomes large. This is a necessary condition for the theory to have a weakly-coupled gravity dual through AdS/CFT, and in which the phenomena of thermalization is related to the black hole formation and evaporation \cite{Hawking:1975IP,Hawking:1976IP}. In fact the two thermodynamic limits can be taken simultaneously, which is then dual in the gravity side to the high temperature ($L \gg \beta$) black holes.  

Studying ETH in the context of quantum field theories (QFTs) has revealed deeper aspects of both thermalization and QFTs. In 2d CFTs, we can study states on a circle of circumference $L=2\pi$ with the spatial coordinate $\varphi\in [0, 2\pi]$. The nature of the ETH becomes more subtle in this context due to the infinite number of symmetry generators forming the Virasoro algebra. Such an algebra gives rise to an infinite number of mutually commuting conserved charges called the KdV charges \cite{Bazhanov:1994KdV,Bazhanov:1996KdV,Bazhanov:1998KdV}. They are constructed from the stress tensor operator. The first few charges are given by: 
\bea
\hat{\mathcal{Q}}_{1}(T) = \int^{2\pi}_0 \frac{d\varphi}{2\pi} T,\;\; \hat{\mathcal{Q}}_{3}(T) = \int^{2\pi}_0 \frac{d\varphi}{2\pi} (TT),\;\;\hat{\mathcal{Q}}_{5}(T) = \int^{2\pi}_0 \frac{d\varphi}{2\pi} \left(T(TT)+\frac{c+2}{12}(\partial T)^2\right)\nonumber
\eea
These charges are universally present and as a result the energy eigenstates are attached with an infinite number of additional labels. The nature of ETH in this context is modified, it is believed that the ``target" equilibrium state corresponds to the so-called generalized Gibbs ensemble (GGE) \cite{Cardy:2016GGE}:
\be \label{eq:GGE_canonical}
\rho_{GGE}(\beta,\tilde{\mu}_i) = \mathcal{N}^{-1} e^{-\beta H(\tilde{\mu}_i)},\;\; H(\tilde{\mu}_i) = \sum_{k\geq 1}\tilde{\mu}_{2k-1} \hat{Q}_{2k-1}(T)
\ee 
where $\mathcal{N}$ is a normalization constant. As a result, the study of subsystem ETH in 2d CFTs involves comparing the entanglement structure of energy eigenstates and those of the equilibrium states such as the GGEs. The simplest eigenstates in 2d CFTs consist of the primary states, which are created via the state-operator correspondence by local primary operators $\mathcal{O}_h$ acting on the vacuum $|\Omega\rangle$ on the complex plane $\mathds{C}$: 
\be\label{eq:radial_q}
|h\rangle = \lim_{x\to 0}\mathcal{O}_h(x)|\Omega\rangle 
\ee  
Properties of these states are computationally the most straightforward to probe. Their relations to thermalization has been studied in \cite{Fitzpatrick:2014,Chen:2017,Wang:2018}; and those to subsystem ETH, e.g. entanglement entropy and Renyi entropies, have been studied \cite{Ryu:2006, Hartman:2013, Faulkner:2013, Hartnoll:2013, Asplund:2015, BinChen:2013, Perlmutter:2014, Lin:2016, He:2017, He:2017p2}. 

In order to study or verify subsystem ETH in 2d CFTs, it is also necessary to reveal the entanglement structures in the thermal equilibrium side. In general, significant entanglement data, e.g. the entanglement spectrum, can be recovered from the knowledge of the Renyi entropies $S_n$ for arbitrary Renyi index $n$. In a companion paper \cite{KdVETHgeneral}, we compute the subsystem entropies for various states in general chaotic CFTs, by assuming certain chaotic ansatz concerning the structure of eigenstate at high charge densities.  In this paper, we focus the computation on the context of AdS/CFT, i.e. we compute the holographic Renyi entropies in thermal equilibrium states of the 2d CFTs. We focus on subsystems that are single intervals on the circle.  

We make some remarks regarding the nature of the equilibrium states considered in this paper. Similar to the distinctions between canonical/micro-canonical ensembles in terms of the conditions imposed on the temperature/energy, with the additional KdV charges one could consider either the GGE represented by (\ref{eq:GGE_canonical}); or the micro-canonical version, i.e. fixing the KdV charges instead of the chemical potential. Although a possibly more appropriate term along the line of GGE should be the ``KdV micro-canonical ensemble", we will refer to the latter simply as the micro-canonical ensemble in this paper. Their density matrices take the form of projection operators on the full Hilbert-space: 
\be 
\rho^{micro}_{q_{2k-1}} = \mathcal{N}^{-1}\hat{P}_{\langle\hat{Q}_{2k-1}\rangle = q_{2k-1}}
\ee 
In the thermodynamic limit, the canonical and micro-canonical ensembles are often considered to be equivalent. However, the equivalence indeed depends on the choice of observables. In particular, it fails for observables that scale exponentially with the large thermodynamic parameter -- when computing expectation values using the saddle point approximation, their ``back-reaction" will cause the two ensembles to differ. Examples of such phenomena include \cite{Wang:2018,Dong:2018}. In the limit of $c\gg 1$ in 2d CFTs, they include heavy operators whose conformal dimension scales with $c$, e.g. the twist operators $\sigma_n$ that compute the Renyi entropies $S_n$ for $n>1$, whose conformal dimensions are given by: 
\be
h_n = \frac{c}{24}\left(\frac{n^2-1}{n}\right) 
\ee
For this reason, in this paper we emphasize the micro-canonical nature of the equilibrium state that appears in the proposal of ETH. The holographic Renyi entropies are computed in the micro-canonical ensembles. For reasons to be explained, we also compute Renyi entropies in more general forms of ensembles with fixed KdV charges. 

In practice holographic computations in these ensembles become more difficult, because the corresponding boundary conditions are less transparent in terms of bulk geometries. To make progress, we will use a scheme of approximation to be introduced in later sections. They work for computing the leading order results in such ensembles with high charge densities. So let us clarify the limits we are working with explicitly. We begin with the $c\to \infty$ scaling ansatz for the CFT chemical potentials $\tilde{\mu}_{2k-1}$:
\be \label{eq:scaling_1}
\tilde{\mu}_{2k-1} = \left(\frac{c}{24}\right)^{-k+1}\mu_{2k-1} 
\ee 
Under such a scaling, the leading order terms in the CFT Hamiltonian (\ref{eq:GGE_canonical}) describe a ``classical" theory of the form: 
\be 
H(\tilde{\mu}_i) = \frac{c}{24} \mathcal{H}(\mu_i)+ \mathcal{O}(c^0),\;\;\mathcal{H}(\vec{\mu}) = \sum_{k} \mu_{2k-1} \mathcal{Q}_{2k-1}(u)
\ee
where the classical density $u$ is related to the CFT stress tensor by:
\be\label{eq:scaling_2}
u(\varphi) = \frac{24}{c} T(\varphi)
\ee 
and $\mathcal{Q}_{2k-1}(u)$ as functions of $u$ are the classical KdV charges, the first few of which are given by:
\be 
\mathcal{Q}_1 (u) = \int^{2\pi}_0 \frac{d\varphi}{2\pi}\; u(\varphi),\;\; \mathcal{Q}_3(u) = \int^{2\pi}_0 \frac{d\varphi}{2\pi}\; u(\varphi)^2,\;\; \mathcal{Q}_5(u) = \int^{2\pi}_0 \frac{d\varphi}{2\pi}\left(u(\varphi)^2 + 2 u'(\varphi)^2\right)\nonumber
\ee
They are related to the quantum KdV charges $\hat{Q}_{2k-1}$ via the rescaling: 
\be\label{eq:scaling_3}
\mathcal{Q}_{2k-1}\sim \left(\frac{c}{24}\right)^{-k} \hat{Q}_{2k-1}
\ee
and taking the leading order part in $c\to \infty$. The ``classical" variables $\lbrace u(x), \mu_{2k-1},\mathcal{Q}_{2k-1}\rbrace$ are what directly enter the holographic calculations. In this paper we will work with them in the context of AdS/CFT; and use (\ref{eq:scaling_1},\ref{eq:scaling_2},\ref{eq:scaling_3}) to convert to the original CFT parameters $\lbrace T(x),\tilde{\mu}_{2k-1} ,\hat{Q}_{2k-1}\rbrace$ when needed. 

On top of these, we are then interested in the limit $\mathcal{Q}_{2k-1} \gg 1$. We shall call this the high charge density limit. Strictly speaking, when defining a sensible micro-canonical ensemble the charges should be allowed to vary in a range of width $\Delta \hat{Q}_{2k-1}$. In this work we take these widths to all be subleading $\Delta \hat{Q}_{2k-1} \ll c^k$, the classical charges $\mathcal{Q}_{2k-1}$ are therefore fixed in the $c\to \infty$ limit of our interest. We can also restore the $L$-dependence by rescaling the spatial coordinates, then the limit corresponds for general circumference $L$ to: 
\be \label{eq:high_density}
\mathcal{Q}_{2k-1} \gg L^{1-2k}
\ee
In terms of the radial quantization states (\ref{eq:radial_q}), we have that: 
\be
\langle \mathcal{Q}_{2k-1} \rangle_h \sim \left(\frac{h}{c}\right)^{k} L^{1-2k}
\ee
Therefore (\ref{eq:high_density}) is satisfied if the ensemble is dominated by the contribution from states $|h \rangle$ satisfying: 
\be
h/c\gg 1,\;\;c\to \infty
\ee
independent of $L$, i.e. for different $L$ the limit (\ref{eq:high_density}) probes parametrically the same regime of the Hilbert space. Having clarified this, from now on we will ignore the $L$-dependence by setting $L=2\pi$ whenever convenient -- especially during explicit computations; and will restore it via dimensional analysis when needed -- mostly for the purpose of stating parametric limits.  

This paper is organized as follows. In section (\ref{sec:KdV}) we first review the basics of the black holes solutions in $\text{AdS}_3/\text{CFT}_2$ that carry KdV charges; we will focus on the BTZ and one-zone black holes that are relevant for latter analysis, and conduct a thorough analysis of their thermodynamic properties in various types of GGEs. In section (\ref{sec:renyi}) we review the holographic computation of Renyi entropies via cosmic-brane backreaction; we introduce a scheme of constructing approximate solutions for the back-reaction called the gluing construction, which works in the high density limit and was first proposed in \cite{Dong:2018}; we then discuss its extension to include higher KdV charges. In section (\ref{sec:main}) we explicitly perform the computation of holographic Renyi entropies in ensembles that fixes the first two KdV charges; we also discuss the implications of the results for the underlying entanglement spectrum. We conclude the paper in section (\ref{sec:discussion}) with some further comments and discussions.

\section{KdV-charged black holes}\label{sec:KdV}
In the holographic (large $c$) limit the gravity background dual to a 2d CFT KdV GGE
\bea
\rho\propto e^{-{\hat{\cal H}}},\qquad \hat{\mathcal{H}}=\sum_{i=}^{m+1} \mu_{2i-1}\hat{\mathcal{Q}}_{2i-1},
\eea 
is a KdV-charged black hole (more carefully, an ensemble of such black holes) \cite{Dymarsky:2020}, with the 3d metric  specified in terms of two functions $f$ and $ u$,
\be\label{eq:finite_zone_metric}
ds^2 = -(fr-\frac{1}{4r}(uf-2f''))^2dt^2+(r+\frac{1}{4r}u)^2d\varphi^2+\frac{dr^2}{r^2}.
\ee
The information about generalized chemical potentials $\mu_{2i-1}$ is encoded in the functional relation between $f$ and $u$ \cite{Perez:2016},
\be 
\label{Hclass}
f[u] = 2\pi\frac{\delta \mathcal{H}(u)}{\delta u},\quad \mathcal{H}(u)=\sum^{m+1}_{i=1} \mu_{2i-1}\mathcal{Q}_{2i-1}(u),\quad {\cal D}f=0,
\ee
where ${\cal D}=\partial_\varphi^3 +u \partial_\varphi$. 
Assuming  number of terms in $\cal H$ is finite, 
the task of finding the black hole solution, i.e.~the function $u(\varphi), 0\leq \varphi\leq 2\pi$ such that $f[u]$ satisfies ${\cal D}f=0$ amounts to finding the so-called finite-zone solution \cite{Novikov:1974}
\bea
\{{\cal H},u\}=0, \label{eq:KdVstatic}
\eea 
with the properly defined Poisson brackets.
To be self-contained, we briefly summarize the procedure below.

\subsection{Finite-zone solutions: a quick review}
We begin by considering the eigenvalue problem for the Schr$\ddot{\text{o}}$dinger equation:
\be\label{eq:schrodinger} 
\Psi^{''}(\varphi) + u(\varphi) \Psi(\varphi) = \lambda \Psi(\varphi),
\ee
i.e. the function $u$ now enters as a periodic potential. For (\ref{eq:schrodinger}) defined on a circle, the discrete spectrum $\lbrace\lambda_n\rbrace$ is defined by requiring periodic/anti-periodic boundary conditions for  $\Psi$,
\be
\Psi(\varphi+2\pi)=\pm \Psi(\varphi) 
\ee
The relation between the Schr$\ddot{\text{o}}$dinger equation (\ref{eq:schrodinger}) and the original KdV equation (whose integrals of motion are the KdV charges) can be understood as follows. The solutions $u(t,\varphi)$ of higher KdV equations
\be\label{eq:KdV} 
\partial_t u = \left\lbrace \mathcal{H}(u),u\right\rbrace,\qquad  \mathcal{H}(u)=\sum_{i=1}^{m+1} \mu_{2i-1}\mathcal{Q}_{2i-1}(u) \
\ee
define the spectrum $\lambda_n(t)$, which a priori is $t$-dependent. But in fact $\lambda_n$ are the integrals of motion, i.e.~$\dot{\lambda}_n(t) = 0$. We are interested in static, $t$-independent solutions satisfying (\ref{eq:KdVstatic}).
These are the so-called finite-zone solution  that have the spectrum $\lbrace \lambda_n\rbrace$ with  all but at most $2m+1$  eigenvalues forming degenerate pairs. 
The subset of non-degenerate eigenvalues $\lbrace \lambda_0<\lambda_1<...<\lambda_{2m}\rbrace$ (together with the information about which zones they correspond to, see below) completely characterizes the  family of static  solutions of \eqref{eq:KdV}. We note that only $m+1$ of these parameters are free, other $m$ parameters are dependent. In addition to $m+1$ independent $\lambda_n$, there are also $m$ parameters which deform $u(\varphi)$ without changing the spectrum -- these are the isospectral deformations generated by first $m$ KdV generators $\mathcal{Q}_{2i-1}$. Thus, in total, the space of $m$-zone solutions is parametrized by $2m+1$ continuous parameters.

These $2m+1$ parameters can be understood as follows. A periodic potential $u(\varphi)$ is an element of the co-adjoint orbit of Witt (Virasoro) algebra. One of these parameters specifies the orbit invariant $h$, related to the monodromy of \eqref{eq:schrodinger} for  $\lambda=0$. Other $2m$ parameters are the coordinates on the symplectic space (reduction of the co-adjoint orbit), with $m$ parameters being the ``action'' variables $I_k$ and other $m$ parameters -- the ``angles'' $\phi_k$. Values of the first $m+1$  KdV charges are the functions of $h$ and $I_k$ (and independent of angles). 
For example, when $m=0$, there is a one-parametric family of  constant solutions $u(\varphi)=Q_1$, parametrized by $Q_1$. When $m=1$, there is a three-parametric family of solutions $u(\varphi-\phi)$, parametrized by $Q_1,Q_3$ and $\phi$.

In addition to continuous parameters, there is $m$ discrete natural numbers $k_{i+1}>k_i$ which specify which zone $\lambda_{2i-1}, \lambda_{2i}$ correspond to. Thus, in the one-zone example above,  the full space of solutions is parametrized by $Q_1,Q_3$, $\phi$ and a positive  integer $k$, as we discuss below.

The $2m+1$ continuous parameters (say, $Q_{2i-1}$ for $1\leq i\leq m+1$ and angles $\phi_k$), together with  $m$ positive integers $k_i$, define the $m$-zone solution $u(\varphi)$, but not $f(\varphi)$. Function $f$, satisfying ${\cal D}f=0$ is mathematically defined only up to an overall coefficient (this implicitly assumes $f$ is sign-definite). One can see an $m$-zone solution as a degenerate $m'$-zone solution with $m'>m$, coming from $\{{\cal H}',u\}=0$ with a different ${\cal H}'$. Accordingly both $f=2\pi \delta {\cal H}/\delta u$ and $f'=2\pi \delta {\cal H}'/\delta u$ will satisfy ${\cal D}f={\cal D}f'=0$ but in general $f\neq f'$. 

Out of $2m+1$ continuous parameters of an $m$-zone $u(\varphi)$,  $m$ parameters can be related to values of $\mu_{2i-1}/\mu_{2m+1}$ in \eqref{eq:KdV}. Thus, if $\mu_{2i-1}$ for $1\leq i\leq m+1$ are specified, the space of static solution is parametrized by $m+1$ additional continuous variables ($h$ and the angles).   Instead of $h$, one can introduce inverse black hole temperature as follows. 
Assuming $f(\varphi)$ is sign-definite (for sign-indefinite $f(\varphi)$ the bulk geometry has the event-horizon stretching all the way to asymptotic boundary, which suggests this case is unphysical), 
one can read out the Bekenstein-Hawking entropy (the horizon area) from the metric
\be\label{eq:finite_zone_entropy}
S = \frac{ \pi \sqrt{u_0}}{2G_N} = \frac{\pi c}{3} \sqrt{u_0}.
\ee
Here $u_0$ is defined as follows
\be \label{eq:entropy_density}
u_0 = \frac{u f^2+f^{'2}-2ff^{''}}{f_0^2},\;\;f_0^{-1} =\frac{1}{2\pi}\int^{2\pi}_0 \frac{d\varphi}{f(\varphi)}.
\ee
The parameter $u_0\equiv h$ labels the co-adjoint orbit $u(\varphi)$ belongs to. For the BTZ solution $u_0=\mathcal{Q}_1$ and (\ref{eq:finite_zone_entropy}) is simply the 2d CFT density of states given by Cardy formula. For a generic finite-zone solution $u_0$ is not the same as the average value of $u(\varphi)$, and thus is not equal to $\mathcal{Q}_1$. 

For a generic finite-zone solution $u$, accompanied by $f$, it can be shown that the numerator of (\ref{eq:entropy_density}) is in fact a constant, and equals to temperature squared \cite{Dymarsky:2020}
\be\label{eq:temp_generic}
(2\pi T)^2 = u f^2+f^{'2}-2ff^{''}, 
\ee
while the sign of $T$ is the sign of $f$.
This fixes $h$ in terms of $\mu_{2m+1}$. Thus, $m+1$ coefficients $\mu_{2i-1}$ fix all continuous parameters of $m$-zone solutions, except for angles.

\subsection{Example: one-zone black hole solutions}\label{sec:one-zone}
As an example  relevant for latter analysis, we examine in detail the one-zone solutions, specified by zone end-points $\lbrace \lambda_1\leq\lambda_2\leq \lambda_3\rbrace$ \footnote{From this section on, we will shift the zone parameter label by 1, i.e. the first zone parameter is $\lambda_1$.}.  These three parameters have to satisfy 
\bea
k = -\pi\frac{\sqrt{\lambda_3-\lambda_1}}{K(p)},\qquad 
p&=& \frac{\lambda_3-\lambda_2}{\lambda_3-\lambda_1},
\eea
where $K(p)$ is the elliptic K function and $k\geq 1$ is a positive integer labeling the zone. Thus the one-zone solutions are parametrized by two continuous and one discrete parameters, in addition to a constant shift of the argument $\varphi$. 

One can choose instead the continuous parameters to be ${\cal Q}_1$ and ${\cal Q}_3$, and the discrete parameter $p$, 
\bea\label{eq:one_zone_parameters}
\mathcal{Q}_1 &=&  4\lambda_3 -4(\lambda_2-\lambda_1)\left(\frac{2\Pi(p,p)}{K(p)}-1\right),\;\;\mathcal{Q}_3 =\frac{J_1-\mu_1 \mathcal{Q}_1}{3}\\ \nonumber
&&J_1 = 16\left(\lambda_1^2+\lambda_2^2+\lambda_3^2-2\lambda_3\lambda_1-2\lambda_1\lambda_2 -2\lambda_2 \lambda_3\right),\\
&& \mu_1 = -8(\lambda_1+\lambda_2+\lambda_3). \nonumber
\eea
where $\Pi(p,p)$ is the complete elliptic integral of the third kind (EllipticPi in Mathematica). Alternatively, it is often convenient to keep $k$ as a discrete parameter, while $p$ will become a continuous parameter together with ${\cal Q}_1$:
\bea\label{eq:one_zone_solution}
\nonumber
&&u(\varphi) = -4\partial^2 \log{\theta\left(ik (\varphi -\phi),q\right)}+\mathcal{Q}_1,\\
&& \theta(ik\varphi,q) = \sum_{n} q^{n^2} \cos{(nk \varphi)},\quad 
q= e^{-\frac{\pi K(1-p)}{K(p)}},\\
&&{\cal Q}_3={\cal Q}_1^2+\frac{64 k^4 (p-1) K(p)^2 \left(K(p)^2+2 (p-2) K(p) \Pi (p, p)-3 (p-1) \Pi (p,p)^2\right)}{3 \pi ^4}.\nonumber
\eea 
Qualitatively, the solution $u(\varphi)$ oscillates along the circle with the frequency that is multiple of $k$, and with the amplitudes controlled by $q(p)$. 
The orbit invariant \eqref{eq:entropy_density} associated with the one-zone solution can be evaluated explicitly, 
\be
\sqrt{u_0} = \left(\sqrt{\frac{4\lambda_1\lambda_2}{\lambda_3}}\right) \frac{\Pi\left(1-\frac{\lambda_2}{\lambda_3},p\right)}{K\left(p\right)}.
\ee

The one zone-solutions span the space of static solutions of \eqref{eq:KdV} for the Hamiltonian of the form
\be\label{eq:GGE_2KdV}
\mathcal{H} = \mathcal{Q}_3 + \mu_1 \mathcal{Q}_1\, . 
\ee 
Without loss of generality we have normalized the coefficient of the $\mathcal{Q}_{3}$ to be one. For a given GGE  $\rho \propto e^{-\beta {\cal H}}$ and a general  static one-zone solution, the zone-parameters are related to the ensemble parameters as follows
\be\label{eq:one_zone_GGE}
2\pi/\beta = 32\sqrt{\lambda_1\lambda_2\lambda_3},\;\;\mu_1-1 = -8(\lambda_1+\lambda_2+\lambda_3),\quad  f(\varphi) = 2u(\varphi) + \mu_1.
\ee

A particular finite-zone solution $u(\varphi)$ specifies corresponding $f(\varphi)$ up to an overall constant. Equation  \eqref{eq:KdV} provides a functional relation between $u$ and $f$. In case of Hamiltonian \eqref{eq:GGE_2KdV} the relation is \eqref{eq:one_zone_GGE}. 
A peculiar feature of the case when $\cal H$ only includes ${\cal Q}_1$ and ${\cal Q}_3$  is that $f(\varphi)$ is always negative, forcing temperature of corresponding black hole background to be negative as well.   
As was pointed out in \cite{Dymarsky:2020}, this means corresponding Euclidean gravitational backgrounds are unstable. They give subleading contribution to the Euclidean path integral dual to GGE state $\rho \propto e^{-\beta {\cal H}}$ with $\cal H$ given by \eqref{eq:GGE_2KdV}, while leading contribution is always given by a BTZ (constant $u(\varphi)$) geometries. 

The same  one-zone solutions $u(\varphi)$  can give rise to a black hole background with positive temperature and even give a dominant contribution to gravitational description of the GGE, when more chemical potentials are turned on \cite{Dymarsky:2020}. For example, let us consider the following GGE, 
\be\label{GGE3}
\rho \propto  e^{-\beta \left(\hat{Q}_5+ \mu_3 \hat{Q}_3 + \mu_1 \hat{Q}_1\right)}.
\ee
Generic black holes in this case are described by two-zone solutions, parametrized by zone-parameters $(\lambda_0,...,\lambda_4)$. However the one-zone solutions are also saddle points of the Euclidean path integral associated with this GGE, provided $\mu_i$ satisfy some additional conditions. 
Given the one-zone solution parametrized by $\lambda_1, \lambda_2, \lambda_3$,  the GGE parameters $\lbrace \beta,\mu_3,\mu_1\rbrace$  must satisfy
\bea\label{eq:one_zone_GGE3}
&&\mu_1 + 48\left(\lambda_1^2+\lambda_2^2+\lambda_3^2\right) + 32\left(\lambda_1\lambda_2+\lambda_2\lambda_3+\lambda_1\lambda_3\right)+8\mu_3\left(\lambda_1+\lambda_2+\lambda_3\right) = 0, \nonumber\\
&&\pi + 16\beta \sqrt{\lambda_1\lambda_2\lambda_3}\left(4\lambda_1+4\lambda_2+4\lambda_3+\mu_3\right)= 0.
\eea

We remark that there are only two equations relating $\lbrace \beta,\mu_3,\mu_1\rbrace$ to $\lambda_i$.  Thus, for fixed $\lbrace \lambda_1,\lambda_2,\lambda_3\rbrace$, one of the GGE parameters, e.g. $\beta$, remains arbitrary, while two others are fixed in terms of $\lambda_i$ and $\beta$. 

\subsection{Smoothness and physical conditions}
For later convenience, we gather here the list of conditions for the one-zone black holes to be physical and smooth as bulk geometries in the GGE (\ref{GGE3}). Related discussions has been performed in \cite{Dymarsky:2020}, for which the details of the derivations can be referred to. We summarize them in the form of inequalities relating the zone parameters $(\lambda_1\leq \lambda_2\leq \lambda_3)$ and the GGE parameters $(\beta,\mu_3,\mu_1)$. Physically these conditions come from the following considerations: 
\begin{itemize}
\item The function $f(\varphi)$ is sign definite, i.e. does not contain zeros. 
\item The temperature $T$ is positive. 
\item The variational response satisfies the first law of thermodynamics with the correct sign:  $\frac{c}{12}d\mathcal{H} = T dS$.
\item The singularities of the metric (\ref{eq:finite_zone_metric}) are covered by the horizon: 
\be
r_H(\varphi)>\text{max}\lbrace r_s(\varphi),0\rbrace,\;\; r_s(\varphi)=-\frac{u}{4},\;\; r_H(\varphi) = \frac{u f^2-2f^{''}f}{4f^2}
\ee
\end{itemize}
These conditions are generic for all finite-zone black hole solutions. It can be checked that they amount to requiring that: 
\be
\forall\varphi\in[0,2\pi],\; f(\varphi)>0;\;\;\; u_0>0;\;\;\;u f^2>f^{''}f
\ee
As remarked before, for a fixed set of zone-parameters $(\lambda_1\leq \lambda_2\leq \lambda_3)$ the corresponding GGE is determined up to a free parameter, which for convenience of the present discussion we choose to be $\mu_3$. In terms of these parameters, the smoothness and physical conditions can be translated into:
\bea \label{eq:smoothness}
&&\lambda_3\geq\lambda_2\geq\lambda_1>0\nonumber\\
&&\mu_3<-4(\lambda_1+\lambda_2+\lambda_3)\nonumber\\
&&\lambda_1\lambda_2+\lambda_1\lambda_3+\lambda_2\lambda_3-\lambda_3^2>0.
\eea 
We make some remarks relating these conditions to the classification of the BTZ black holes, i.e. whether they are deformable or isolated. As discussed before, in the limit of coincident zone-parameters, i.e. $\lambda_1=\lambda_2=w/4$ or $\lambda_2=\lambda_3=w/4$, the one-zone black hole reduces to a BTZ black hole. From this we can infer that a BTZ black hole is deformable if its zone-parameters $(h,w)$ satisfy the smoothness and physical conditions (\ref{eq:smoothness}); otherwise it is isolated. For isolated BTZ black holes, the only condition is the positivity of the mass, i.e. $h=\langle \mathcal{Q}_1\rangle>0$. There is no restriction for the remaining parameter $w$ -- it could even be complex. The arguments leading to (\ref{eq:smoothness}) assume real zone-parameters to begin with, which is indeed necessary for non-degenerate one-zone black holes. On the other hand, for those isolated BTZ black holes with real $w<0$, it is interesting to understand how do they as smooth solutions evade the arguments leading to (\ref{eq:smoothness}). We provide some details discussing this in the appendix (\ref{sec:smooth_BTZ}).

\subsection{Thermodynamics}\label{sec:thermo_onezone}
In this subsection, we analyze the thermodynamic properties of these one-zone black holes in the context of GGE (\ref{GGE3}) with 3 KdV chemical potentials:
\be
\rho= \mathcal{N}^{-1} e^{-\beta\mathcal{H}},\;\;\mathcal{H}=\hat{Q}_5+\mu_3 \hat{Q}_3 +\mu_1 \hat{Q}_1  
\ee

\subsubsection{Phases of BTZ solutions}
We are in particular interested in the thermodynamics of the one-zone black holes that are perturbatively close to a BTZ solution in the GGE (\ref{GGE3}). To this end, we first study the properties of BTZ solutions, in particular how do they depend on the BTZ parameters $(h,w)$ as well as the GGE parameters $(\beta,\mu_1,\mu_2)$ they are in. Recall that $(h,w)$ is related to $(\beta,\mu_1,\mu_3)$ by restricting (\ref{eq:one_zone_GGE3}) to cases with two coincident zone parameters:
\bea\label{BTZ_equation}
&&T=\mathcal{G}(h),\;\;\mathcal{G}(h)\equiv\frac{1}{2\pi}(3h^{5/2}+2\mu_3 h^{3/2}+\mu_1 h^{1/2})\nonumber\\
&&8w^2+4(h+\mu_3)w+\frac{2\pi}{\beta \sqrt{h}}=0.
\eea
where $T=1/\beta$ is the temperature. Positive roots to the first equation are identified as the masses of the BTZ black holes in the GGE. It is easily recognized as the saddle-point equation for the primary state contribution to the partition function: 
\be\label{eq:primary_pf}
Z(\beta,\mu_1,\mu_3) \sim \int dh\; e^{-\beta\;\mathcal{F}_{BTZ}(h)},\;\;\mathcal{F}_{BTZ}(h) = \frac{c}{12}(h^3+\mu_3 h^2+\mu_1 h)- S(h)  
\ee
From (\ref{BTZ_equation}), there are only three independent parameters specifying a BTZ black hole together with the GGE. We choose $(h,\beta,\mu_3)$ to facilitate latter discussions.  Notice that at fixed $(h,\beta,\mu_3)$, while $\mu_1$ is uniquely determined, there are two solutions $w^{\pm}$ to the second equation of (\ref{BTZ_equation}). We interpret this as potentially two branches of one-zone black holes whose limits of either $p\to 0$ or $p\to 1$ give rise to the BTZ at the prescribed $(h,\beta,\mu_3)$. The properties that are relevant to us include the following key aspects: 
\begin{itemize}
\item \textbf{Extremum type for $\mathcal{F}_{BTZ}(h):$} positive roots of the first equation in (\ref{BTZ_equation}) are extremum of $\mathcal{F}_{BTZ}(h)$. The BTZ has to be a local minimum of $\mathcal{F}_{BTZ}(h)$ before surviving as the thermodynamically stable saddle of the full GGE. This can be checked by computing $\mathcal{F}^{''}(h)$ at the roots, from which we obtain that a BTZ at $(h,\beta,\mu_3)$ is a local minimum if: 
\be\label{eq:BTZ_localmin}
3h+\mu_3 > - \left(\frac{\pi}{2\beta h^{3/2}}\right) 
\ee
\item \textbf{Deformable v.s. isolated type:} For the BTZs satisfying (\ref{eq:BTZ_localmin}), we are interested in whether they can be deformed into nearby one-zone black holes satisfying (\ref{eq:smoothness}). As discussed before, this depends on whether the BTZ itself satisfies (\ref{eq:smoothness}), which amounts to the following inequalities: 
\be\label{eq:deform_condition}
\mu_3+h+2w<0,\;\; w>\left(\sqrt{2}-1\right)h >0
\ee
It turns out that when $(h,\beta,\mu_3)$ satisfy: 
\bea\label{eq:deform_zone}
h+\mu_3< -\left(\frac{4\pi}{\beta \sqrt{h}}\right)^{1/2}<0
\eea 
Both branches of solutions $w^{\pm}$ are positive: 
\be\label{eq:w_branches}
w^{\pm} = -\left(\frac{h+\mu_3}{4}\right)\left(1\pm\sqrt{1-\frac{4\pi}{\beta \sqrt{h}(h+\mu_3)^2}}\right)>0
\ee
Furthermore, they both satisfy the first inequality in (\ref{eq:deform_condition}):
\be
h+\mu_3+2w^{\pm} = -w^{\mp}<0 
\ee
It is then left to checking the second inequality in (\ref{eq:deform_condition}) to determine whether they are deformable. In particular, if the BTZ corresponds to the $p\to 0$ limit of one-zone black holes, then $w>h>(\sqrt{2}-1)h$ and it is automatically deformable. 

\item \textbf{Limit type of deformable BTZs:} For those deformable BTZ black holes satisfying (\ref{eq:deform_condition}), we are then interested in whether they correspond to the $p\to 0$ limit or the $p\to 1$ limit of one-zone black holes. As mentioned previously this depends on the sign of $\Delta =h-w$, which satisfies the quadratic equation derived from (\ref{BTZ_equation}):
\be
4\Delta^2-2(5h+\mu_3)\Delta +\left(6h^2+2h\mu_3+\frac{\pi}{\beta \sqrt{h}}\right)=0 
\ee 
For local minimum satisfying (\ref{eq:BTZ_localmin}), both branches $\Delta^{\pm}$ are of the same sign as that of $5h+\mu_3$. We therefore conclude that they correspond to the $p\to 0$ limit for both branches $w^{\pm}>h$ if $5h+\mu_3<0$; while for $5h+\mu_3>0$ they correspond to the $p\to 1$ limits for both branches $w^{\pm}<h$. 
\end{itemize}
Based on these, we can derive the following phases regarding the BTZ black hole at fixed $h$ as one vary the remaining parameters: 
\be\label{eq:dimless}
\chi_1 = \left(\frac{\pi T}{h^{5/2}}\right),\;\;\chi_2 = \left(\frac{\mu_3}{h}\right) 
\ee
The phases are organized into windows of $\chi_2$ whose locations as well as structure vary with $\chi_1$, see Table (\ref{tb:phases}). 
\begin{table}[h]
\centering
\hspace{-0.1cm}
\begin{minipage}{0.75\textwidth}
\begin{tabular}{|c|c|}
\hline
\raisebox{3\height}{$0<\chi_1<\infty$ }&\includegraphics[width=0.4\textwidth]{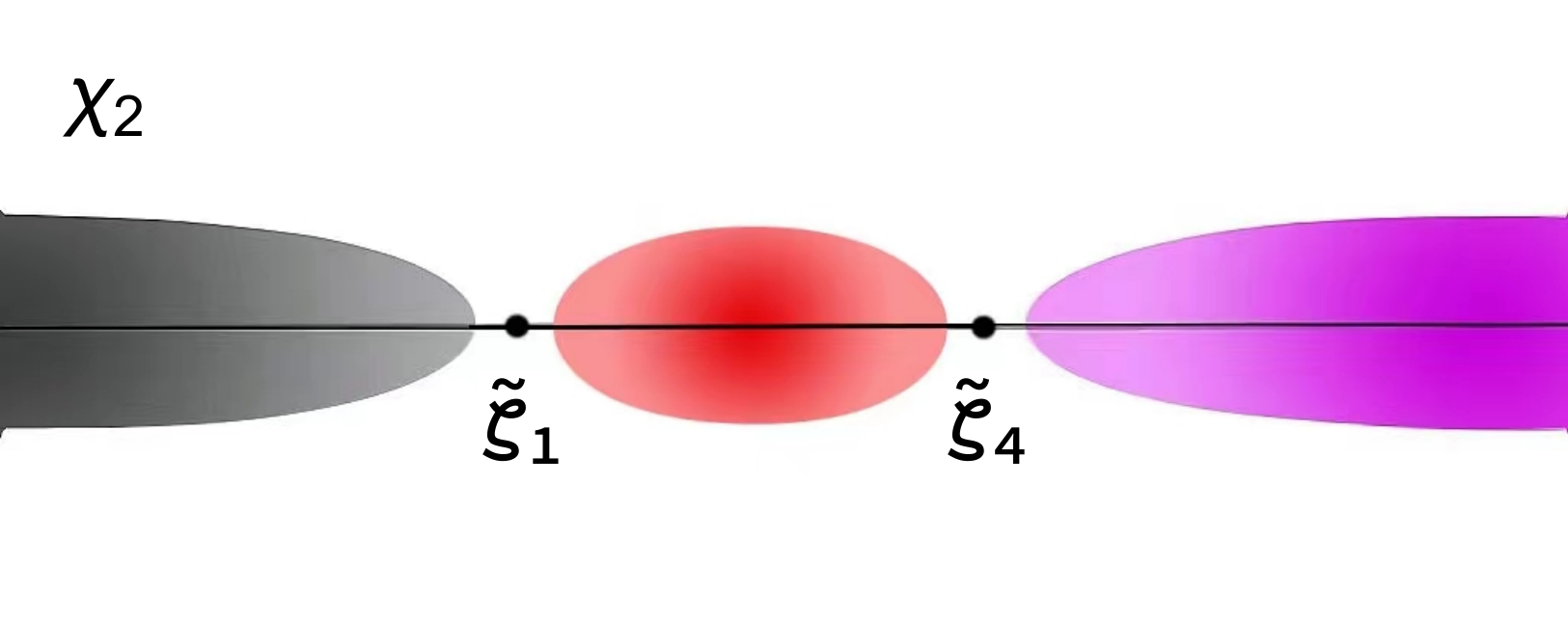} \\
\hline  
 \raisebox{4\height}{$\alpha_2<\chi_1<\alpha_3$} & \includegraphics[width=0.4\textwidth]{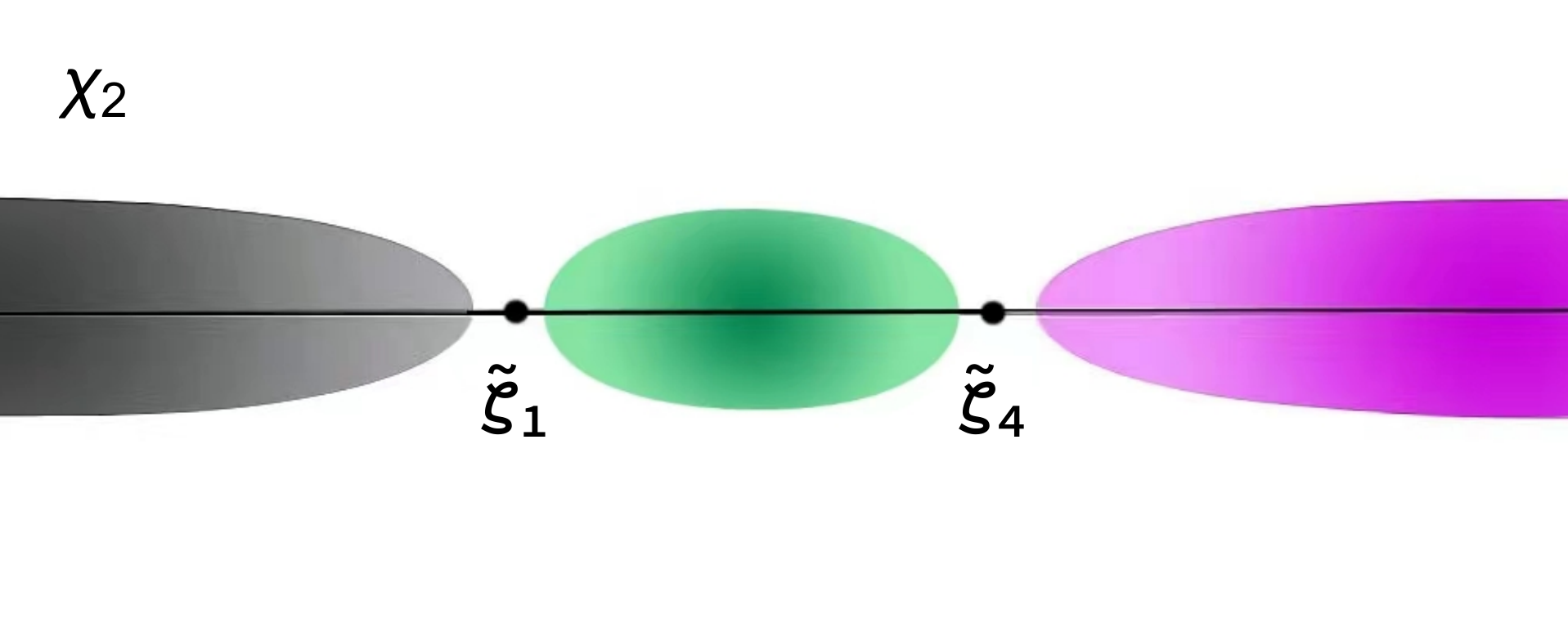} \\
\hline
\raisebox{4\height}{ $\alpha_1<\chi_1<\alpha_2$} & \includegraphics[width=0.4\textwidth]{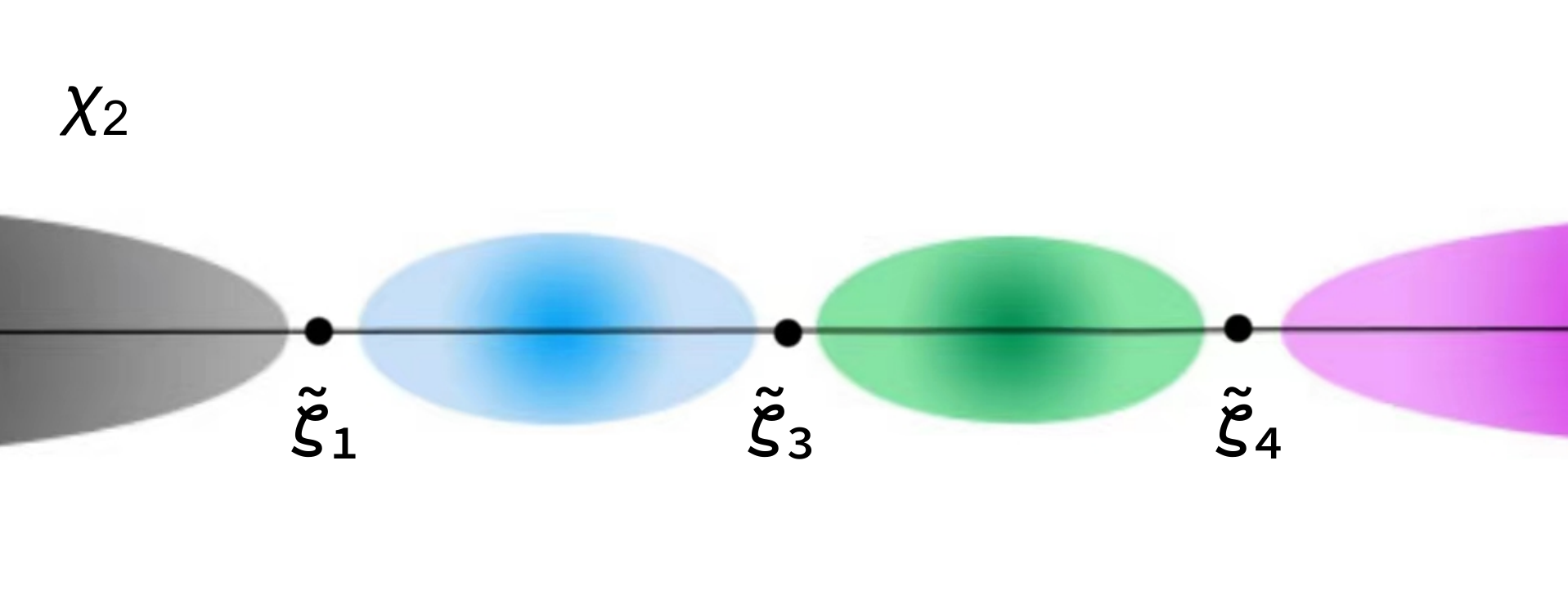} \\
\hline
 \raisebox{3\height}{$0<\chi_1<\alpha_1$} & \includegraphics[width=0.4\textwidth]{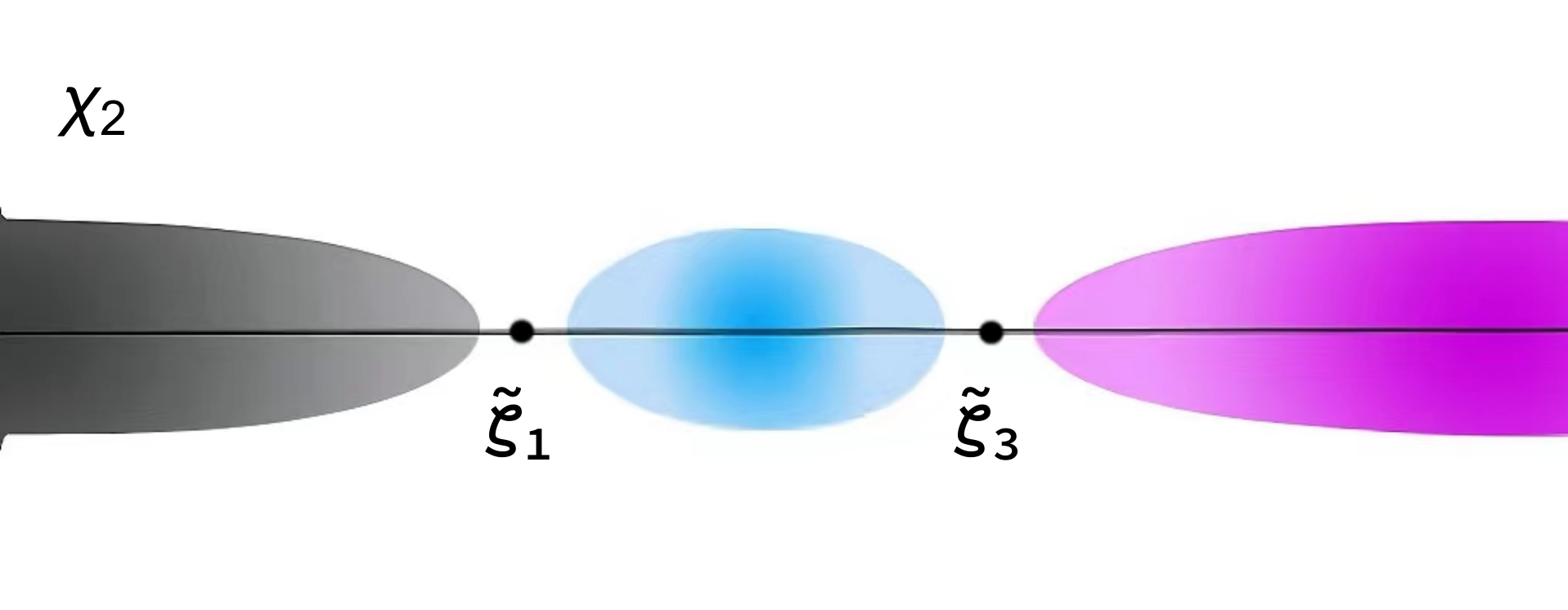} \\
\hline
\end{tabular}
 \caption{The phases of BTZ black holes}\label{tb:phases}
    \end{minipage}
\hspace{-2cm} 
\begin{minipage}{0.35\textwidth}
    \vspace{-1cm}
    \hspace{-1.5cm} 
        \raggedright
   \textcolor{gray}{\rule{1em}{1em}} $h$ is a local maximum of $\mathcal{F}_{BTZ}$\par \vspace{0.2cm}
   \hspace{-1.5cm} 
   \textcolor{red}{\rule{1em}{1em}}   $(h,w^{\pm})$ are deformable to $p\to 0$\par \vspace{0.2cm} \hspace{-1.5cm} 
   \textcolor{teal}{\rule{1em}{1em}} $(h,w^{\pm})$ are deformable to $p\to 1$ \par \vspace{0.2cm} \hspace{-1.5cm} 
   \textcolor{cyan!60}{\rule{1em}{1em}} $(h,w^+)$ is deformable to $p\to 1$ \par \vspace{0.2cm} \hspace{-1.5cm} 
    \textcolor{violet!60}{\rule{1em}{1em}} both $(h,w^{\pm})$ are isolated\par 
    \end{minipage}    
    \end{table}
The ranges of of $(\chi_1,\chi_2)$ are defined by intervals whose boundaries occur at the following values:
\bea
\alpha_1 &=& 4(\sqrt{2}-1)^2,\alpha_2=4(\sqrt{2}-1),\alpha_3=4,\;\;\tilde{\zeta_1}= -\chi_1/2-3\nonumber\\
\tilde{\zeta_2}&=&-5,\tilde{\zeta_3}=-(2\sqrt{2}-1)-\chi_1/(2\sqrt{2}-2),\tilde{\zeta}_4=-2\sqrt{\chi_1}-1 
\eea

More details for the derivation are included in the appendix (\ref{app:ReBTZ}). When the BTZ is deformable, it is likely to be thermodynamically unstable against nearby one-zone black holes in the GGE, one needs to further compute the free energies; when it is isolated we view it as thermodynamically stable, at least locally. 

\begin{figure}[ht]
	\centering
	\includegraphics[width=0.5\linewidth]{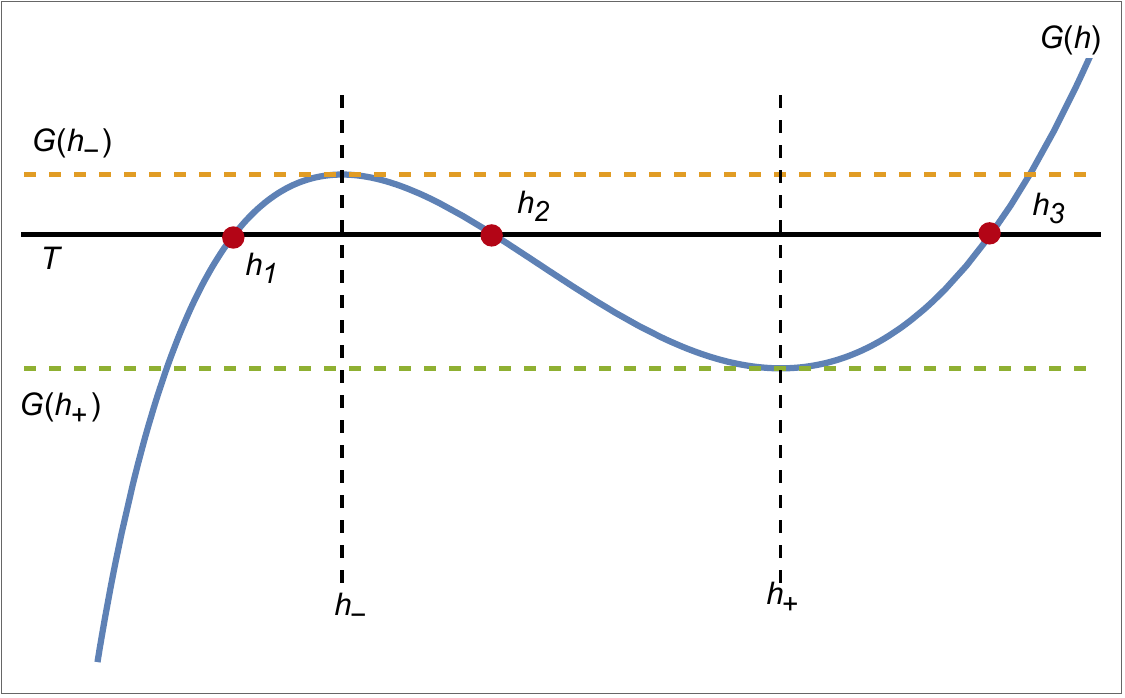}
	\caption{A GGE satisfying (\ref{eq:3_BTZ}) has three positive roots $(h_1<h_2<h_3)$.}
	\label{fig:four_types}
\end{figure}

We can also focus only on the GGE parameters $(\beta,\mu_1,\mu_3)$, and identify a phase where it contains two physical BTZ black holes. This corresponds to when the first equation in (\ref{BTZ_equation}) has three positive roots $(h_1<h_2<h_3)$. It is easy to see that $(h_1,h_3)$ are local minimum and $h_2$ is a local maximum for $\mathcal{F}_{BTZ}(h)$. This can only happen if $\mathcal{G}(h)$ has two positive turning points  $0<h_-<h_+,\;\mathcal{G}'(h_{\pm})=0$, of which $h_-$ is a local maximum and $h_+$ is a local minimum for $\mathcal{G}(h)$; and the positive temperature is between the two extrema: $\mathcal{G}(h_+)<T<\mathcal{G}(h_-)$, see Fig. \eqref{fig:four_types}. This can be translated into the following conditions for $(T,\mu_1,\mu_3)$: 
\bea\label{eq:3_BTZ}
&&\mu_3<0,\;\;\;0<\mu_1<\frac{3}{5}\mu_3^2,\;\;\; \text{max}\lbrace0,\;\mathcal{G}(h_+)\rbrace <T<\mathcal{G}(h_-)\nonumber\\
&& h_{\pm} = -\frac{\mu_3}{5}\left(1\pm \sqrt{1-\frac{5\mu_1}{3\mu_3^2}}\right)
\eea
When the BTZ solutions at $h_{1,3}$ are deformable, it is easy to see that both branches $(h_1,w^{\pm}_1)$ are $p\to 0$ limits; and $(h_3,w^{\pm}_3)$ are $p\to 1$ limits. Among the local minimum $(h_1,h_3)$ of $\mathcal{F}_{BTZ}(h)$, only one of them corresponds to the global minimum and is likely to be thermodynamically stable.

\subsubsection{Thermodynamic stabilities near BTZ}
The perturbative expansion of various quantities in these limits can be computed. Let us label the one-zone solution and the corresponding GGE after solving (\ref{eq:one_zone_GGE3}) by $\lbrace \lambda_1,\lambda_3,p,\mu \rbrace$, the results at leading orders can be summarized below.

We first study the thermodynamic stability near the $p\to 0$ limit. In this limit, the deviations from the BTZ solutions are controlled by powers of $p$. In particular, the KdV charges and the entropy density are given to the leading orders in $p$ by: 
\bea
\langle \mathcal{Q}_1\rangle &=& 4\lambda_1+\frac{1}{2}p^2(\lambda_3-\lambda_1) \nonumber\\
\langle \mathcal{Q}_3\rangle &=& 16\lambda_1^2+4p^2(\lambda_3-\lambda_1)(2\lambda_3-\lambda_1)\nonumber\\
\langle \mathcal{Q}_5\rangle &=&64\lambda_1^3+8p^2(\lambda_3-\lambda_1)(8\lambda_3^2-4\lambda_1\lambda_3-\lambda_1^2)
 \label{eq:p=0_kdv} \\
\sqrt{u_0} &=& 2\sqrt{\lambda_1}+p^2\sqrt{\lambda_1}\frac{\lambda_3-\lambda_1}{8\lambda_3}  \label{eq:p=0_su0}
\eea
However we are more interested in the difference between the BTZ solution and the one-zone solution when the GGE is specified. In particular, we care about how the free energy changes when we deform a BTZ solution. In the zero-zone limit, the zone parameters can be solved perturbatively in terms of the GGE parameters
according to \eqref{eq:one_zone_GGE3}, therefore the free energy difference can be obtained. We find that the difference $\delta F=F_{\text{one-zone}}-F_{\text{BTZ}}$ starts to show up in the order of $p^4$:
\bea 
\delta F = \frac{p^4h^3(\tilde{w}-1)^3}{64 \tilde{w}(\chi_1-4\tilde{w})}\left(60\tilde{w}^4-7\chi_1 \tilde{w}^3-19\chi_1 \tilde{w}^2+\chi_1\left(\frac{3}{4}\chi_1-4\right)\tilde{w}+3\chi_1^2\right)
\eea 
where we have written the result in terms of $(\chi_1,\chi_2)$ defined in (\ref{eq:dimless}) and $\tilde{w}=w/h$. Recall from (\ref{eq:w_branches}) that for the BTZ parametrized by $(h,T,\mu_3)$ it has two branches: 
\be\label{eq:wt_branches}
\tilde{w}^{\pm} = -\left(\frac{1+\chi_2}{4}\right)\left(1\pm \sqrt{1-\frac{4\chi_1}{(1+\chi_2)^2}}\right)
\ee
For the BTZ to be deformable consistently as the $p\to 0$ limit, the parameters $(\chi_1,\chi_2)$ can only be in the window: 
\be\label{eq:p0BTZ}
\chi_1>4,\;\;\;\tilde{\zeta}_1<\chi_2<\tilde{\zeta}_4
\ee 
Within this range $\tilde{w}^{\pm}$ are constrained to vary in the intervals: 
\be
1< \tilde{w}^- <  \frac{\sqrt{\chi_1}}{2} <\tilde{w}^+ < \frac{\chi_1}{4}
\ee
One can check that at fixed $\chi_1>4$, the free energy cost $\delta F$ as a function of $\tilde{w}$ satisfies: 
\be\label{eq:dF_sign}
\delta F\left(\frac{\chi_1}{4}\right)>0,\;\;\delta F\left(\frac{\sqrt{\chi_1}}{2}\right)<0,\;\;\delta F(1)>0
\ee 
This implies that for both branches $w^{\pm}$, there must be a phase transition regarding the sign of $\delta F$ within the range (\ref{eq:p0BTZ}). More specifically, the branches $w^{\pm}$ are thermodynamically \textbf{unstable} against nearby one-zone black holes for $\left(y_{\pm}<\chi_2<\tilde{\zeta}_4\right)$; and are \textbf{stable} for $\left(\tilde{\zeta}_1<\chi_2<y_{\pm}\right)$. The threshold values $y_{\pm}$ are the roots of the following equations: 
\be\label{eq:thresholds}
1\pm \sqrt{1-\frac{4\chi_1}{(1+y_{\pm})^2}} = \frac{60\chi_1(y_{\pm}+1)^2+2\chi_1^2(7y_{\pm}-85)}{(y_{\pm}+1)\left(30(1+y_{\pm})^3-10\chi_1^2-\chi_1(75+84y_{\pm}-7y_{\pm}^2)\right)}\nonumber
\ee 
that are constrained to lie in: 
\be 
\tilde{\zeta}_1 < y_{\pm} < \tilde{\zeta}_4
\ee
They are guaranteed to exist due to (\ref{eq:dF_sign}). 

Next we look at the $p\to 1$ limit. It turns out in this limit, the leading order deviations from the BTZ solution are controlled by $\Lambda^{-1} \ll 1$, where: 
\be
\Lambda = -\ln{\left(\frac{1-p}{16}\right)} \gg 1
\ee  
The KdV charges and entropy density are given to the leading order in $\Lambda^{-1}$ by: 
\bea
\langle\mathcal{Q}_1\rangle &=& 4\lambda_3 - \frac{16}{\Lambda}(\lambda_3-\lambda_1)\nonumber\\
\langle\mathcal{Q}_3\rangle &=& 16\lambda_3^2-\frac{128}{3\Lambda}(\lambda_3-\lambda_1)(2\lambda_1+\lambda_3)\nonumber\\
\langle\mathcal{Q}_5\rangle &=& 64\lambda_3^3-\frac{256}{5\Lambda}(\lambda_3-\lambda_1)(3\lambda_3^2+4\lambda_1\lambda_3+8\lambda_1^2)\label{eq:p=1_kdv}\\
\sqrt{u_0} &=& 2\sqrt{\lambda_3}-\frac{4}{\Lambda}\sqrt{\lambda_3-\lambda_1} \tanh^{-1}\left(\sqrt{1-\frac{\lambda_1}{\lambda_3}}\right)\label{eq:p=1_su0}
\eea
Similarly, after specifying the GGE the difference of the free energy is
\bea \label{eq:Fcost_p1}
\delta F &=& \frac{8h^3}{\Lambda}\Big[\left(\frac{1-\tilde{w}}{15\tilde{w}}\right)\left(8\tilde{w}^3-16\tilde{w}^2+(8-10\chi_1)\tilde{w}-5\chi_1\right)+ \chi_1\sqrt{1-\tilde{w}}\nonumber\\
&\times&\tanh^{-1}\left(\sqrt{1-\tilde{w}}\right)\Big]
\eea 
The BTZ is deformable consistently as the $p=1$ limit of one-zone black holes if the parameters satisfy : 
\be
\chi_1<4,\;\;\;\text{max}\left\lbrace \sqrt{2}-1,\;\frac{\chi_1}{4}\right\rbrace <\tilde{w}^-<\frac{\sqrt{\chi_1}}{2}<\tilde{w}^{+} <1 
\ee

It can be verified that in this range, the branch $w^+$ is always thermodynamically \textbf{stable}, i.e. featuring a positive definite free energy cost $\delta F>0$ to nearby one-zone black holes. However the branch $\tilde{w}^-$, when exists, contains a phase transition -- it becomes thermodynamically \textbf{unstable} for $\chi_2<y$. The threshold value $y$ is the single root of (\ref{eq:Fcost_p1}) for $\chi_2$ through its dependence from plugging $w^-$ in (\ref{eq:wt_branches}). 

\subsubsection{Ensembles at fixed KdV charges: micro-canonical and mixed}\label{sec:mmGGE}
As mentioned in the introduction, it is the micro-canonical ensemble whose KdV charges are fixed that is the most closely related to ETH in 2d CFTs. There are infinitely many KdV charges that one can fix in principle, in this paper we discuss fixing only a finite number. The simplest such ensemble is the those fixing only $\langle \mathcal{Q}_1\rangle$ and $\langle \mathcal{Q}_3 \rangle$:
\be
\rho^{micro}_{q_1,q_3} = \mathcal{N}^{-1} \hat{P}_{q_1,q_3} 
\ee
where $\hat{P}_{q_1,q_3}$ denotes the projector into the Hilbert sub-space with the prescribed KdV charges: 
\be\label{eq:q1q3}
\langle\mathcal{Q}_1\rangle = q_1,\;\;\langle \mathcal{Q}_3\rangle =q_3 
\ee
The first question one naturally asks about the micro-canonical ensemble $\rho^{micro}_{q_1,q_3}$ is whether it has a well-defined bulk dual description. Abstractly, $\rho^{micro}_{q_1,q_2}$ is related to $\rho_{\mu_1,\mu_3}$ by an inverse Laplace transform: 
\bea 
\rho^{micro}_{q_1,q_3} &=& \oint_{\Gamma_1} d\mu_1 \oint_{\Gamma_3} d\mu_3\;e^{\mu_1 q_1+\mu_3 q_3}\rho_{\mu_1,\mu_3}\nonumber\\
\rho_{\mu_1,\mu_3} &=&\mathcal{N}^{-1} e^{-\tilde{\mu}_1 \hat{Q}_1 -\tilde{\mu_3} \hat{Q}_3}
\eea
where $\Gamma_{1,3}$ are the corresponding Bromwich contours for $\mu_{1,3}$. In the thermodynamic limit, the inverse Laplace transform can proceed by simply finding the saddle-points for $\mu_{1,3}$. In physical terms this means finding a particular $(\mu^*_{1,3})$ whose KdV charges $\langle \mathcal{Q}_1\rangle$ and $\langle \mathcal{Q}_3\rangle$ match with their prescribed values $(q_1,q_3)$. In the context of $\text{AdS}_3/\text{CFT}_2$, its bulk dual will be a black hole solution at chemical potentials $\mu^*_1,\mu^*_3$ giving the corresponding KdV charges. For generic values of $q_3\neq q_1^2$, they have to be one-zone black holes. However, at two chemical potentials $(\mu_1, \mu_3)$ only the BTZ black holes are physical Euclidean saddles of the GGE. One therefore infers that at generic fixed KdV charges $q_3 \neq q_1^2$, the micro-canonical ensembles do not admit well-defined bulk duals. 

We have studied one-zone black holes in the GGEs (\ref{eq:GGE_q5}) featuring three chemical potentials, in which they exhibit well-defined thermodynamic properties. Motivated by this, we can consider more general forms of ensembles with fixed $\langle \mathcal{Q}_1\rangle$ and $\langle \mathcal{Q}_3\rangle$. For example, we can consider the following ensembles: 
\be
\rho^\beta_{q_1,q_3} = \mathcal{N}^{-1}\hat{P}_{q_1,q_3}\;e^{-\beta \hat{Q}_5} 
\ee
They describe a non-uniform distribution in the micro-canonical shell of KdV charges $(q_1,q_3)$, the statistical weight is decorated by a temperature associated with $\langle \mathcal{Q}_5\rangle$. We refer to them as the mixed ensembles in this paper. They can be obtained from the GGE (\ref{eq:GGE_q5}) via a similar Laplace transform: 
\bea 
\rho^{\beta}_{q_1,q_3} &=& \oint_{\Gamma_1} d\mu_1 \oint_{\Gamma_3} d\mu_3\;e^{\mu_1 q_1+\mu_3 q_3}\rho_{\beta,\mu_1,\mu_3}\nonumber\\
\rho_{\beta,\mu_1,\mu_3} &=&\mathcal{N}^{-1} e^{-\tilde{\mu}_1 \hat{Q}_1 -\tilde{\mu_3} \hat{Q}_3 - \beta \hat{Q}_5}
\eea 
Similarly, in the thermodynamic limit this is given via the saddle-point approximation by a black hole solution in the GGEs (\ref{eq:GGE_q5}) whose first two KdV charges coincide with the prescribed values $(q_1,q_3)$. This requirement does not uniquely fix the black hole solution. To determine the equilibrium configuration of $\rho^{\beta}_{q_1,q_3}$, we need to find the black hole solution that minimizes the free energy: 
\be\label{eq:F_mmGGE}
F^{\beta}_{q_1,q_3} \propto \mathcal{Q}_5 - T S 
\ee 
For the GGEs (\ref{eq:GGE_q5}) the most generic black holes are two-zone solutions. In this paper, we focus on the one-zone sector. We assume that the two-zone black holes tend to cost higher free energies, though this should be checked more rigorously in future investigations. 

Eliminating two of the three zone parameters $(\lambda_1\leq \lambda_2\leq \lambda_3)$ using the constraint on the fixed KdV charges \eqref{eq:q1q3}, there is one free parameter remain. We take it to be $p=\frac{\lambda_3-\lambda_2}{\lambda_3-\lambda_1}$, which then parametrizes the micro-canonical shell of one-zone black holes.  For each fixed $p$ there are two branches of solutions for the zone-parameters satisfying (\ref{eq:q1q3}). Only one of them corresponds to zone parameters that are likely to be physical: 
\bea\label{eq:zone_mmGGE} 
&&\lambda_1=\frac{q_1}{4} + \frac{\sqrt{3(q_3-q_1^2)} \left((p-2) K(p)+2 E(p)\right)}{8 \sqrt{(p-1) K(p)^2-2(p-2) K(p) E(p)-3 E(p)^2}}\nonumber \\
&&\lambda_3=\frac{q_1}{4} + \frac{\sqrt{3(q_3-q_1^2)} \left((p-1) K(p)+2 E(p)\right)}{8 \sqrt{(p-1) K(p)^2-2(p-2) K(p) E(p)-3 E(p)^2}}\nonumber\\
&&\lambda_2= (1-p ) \lambda_3+ p \lambda_{1}
\eea
It can be checked from (\ref{eq:zone_mmGGE}) that $\lambda_2,\lambda_3 \to +\infty$ in the $p\to 0$ limit; while $\lambda_1,\lambda_2\to -\infty$ in the $p\to 1$ limit. The latter limit does not give physical one-zone black holes. It is therefore important to find the range of $p$ parametrizing the physical one-zone black holes satisfying (\ref{eq:q1q3}). To this end, we plug the zone parameters (\ref{eq:zone_mmGGE}) as functions of $p$ in the smoothness and physical conditions (\ref{eq:smoothness}) and derive the bound on $p$. It turns out that the tightest bound comes from the last condition of (\ref{eq:smoothness}), which prohibits naked singularities in the one-zone black hole geometry: 
\be
\lambda_1\lambda_2+\lambda_1\lambda_3+\lambda_2\lambda_3-\lambda_3^2>0 
\ee
This imposes the following bound on the allowed range of $p$
\be \label{eq:p_range}
0\leq p\leq p_+
\ee
The upper bound $p_+$ is the solution of a transcendental equation, and depends on the fixed KdV charges $(q_1,q_3)$. For the purpose of latter discussions, we are interested in the following limit \footnote{This is slightly different from the parameter $\epsilon=q_3/q_1^2-1$ defined in \cite{KdVETHgeneral}.}: 
\be
\epsilon = \frac{1}{q_1}\sqrt{q_3-q_1^2} \ll 1 
\ee  
In this limit, we can compute the leading order result for $p_+$: 
\be\label{eq:p_bound}
p_+ =  1 - 16\exp(-\Lambda_{+} ), \quad \Lambda_{+} = \frac{32(3-2 \sqrt{2})}{3 \epsilon^2} 
\ee

Now we can discuss the thermodynamics of the mixed ensembles $\rho^\beta_{q_1,q_3}$ based on the allowed one-zone black holes (\ref{eq:zone_mmGGE}) in the range (\ref{eq:p_range}). For a fixed temperature $T=1/\beta$, the equilibrium configuration corresponds to the particular one-zone black hole parametrized by $p_T^*$, which minimizes the free energy (\ref{eq:F_mmGGE}): 
\be 
\frac{\partial F^\beta_{q_1,q_3}}{\partial p}\Big|_{p^*_T}= \left(\frac{\partial \mathcal{Q}_5}{\partial p}-T \frac{\partial S}{\partial p}\right)\Big|_{p^*_T} =0
\ee
Next we discuss the equilibrium value $p^*_T$ as a function of the temperature $T$. It can be checked that $S$ is maximized to be: 
\be
S\to \frac{\pi \sqrt{q_1}}{2G_N} 
\ee 
in the $p\to 0$ limit. On the other hand, $\mathcal{Q}_5 \to \infty$ in the same limit. Such an interplay between the two terms in $F^{\beta}_{q_1,q_3}$ implies that the equilibrium $p^*_T$ admis a high temperature expansion near $p^*_T=0$. At the leading order, it can be computed in terms of the rescaled temperature $\chi_1 = \pi T/q_1^{5/2}$ by: 
\be\label{eq:p_highT}
p^*_T =\epsilon \sqrt{\frac{8}{\chi_1}} +... \;\;\; \chi_1 \gg \epsilon^2
\ee
From this result, we can also obtain a corresponding high temperature expansion of the thermodynamic entropy $S= \pi\sqrt{u_0}/2G_N$ for the mixed ensemble $\rho^{\beta}_{q_1,q_3}$, where: 
\be 
\sqrt{ u_0} = \sqrt{ q_1}\left(1 - \frac{\epsilon^2}{4\sqrt{\chi_1}} +...\right)
\ee
It is interesting to take the infinite temperature limit $T\propto \chi_1 \to \infty$. Doing this recovers the microcanonical ensemble at fixed $(q_1,q_3)$. We discover that the entropy reduces to that of the ordinary microcanonical ensemble at fixed $q_1$ at the leading order in $G_N$: 
\be\label{eq:microKdV_entropy}
S_{micro}(q_1,q_3) = \frac{\pi \sqrt{q_1}}{2G_N}+...
\ee
We clarify some subtleties here. In taking $T\to \infty$, the equilibrium bulk configuration approaches the $p\to 0$ limit of one-zone black holes, which signals the degeneration into a BTZ black hole. On the other hand, the KdV charges of the BTZ black holes at the leading order in 1/c always satisfy: $q_3 = q_1^2 \to \epsilon =0$. This is in contradiction with the charges we are fixing in $\rho^{micro}_{q_1,q_3}$. What happened is that at finite $\epsilon$, the limit $p\to 0$ also drives two of the zone-parameters to diverge: 
\be 
\lambda_2, \lambda_3 \sim q_1\sqrt{\frac{\epsilon^2}{8p}} \to \infty.
\ee
The result seems to suggest that despite not having a Euclidean bulk dual, the micro-canonical ensemble at fixed $(q_1,q_3>q_1^2)$ can be interpreted as a BTZ black hole decorated with a macroscopic condensation of bulk ``hair" that accommodates the surplus $Q_3$ charges. In the companion paper \cite{KdVETHgeneral}, similar results regarding the micro-canonical entropy at fixed $(q_1,q_3)$ at large $c$ are also obtained using more general approaches.

As the temperature lowers, the equilibrium value $p^*_T$ increases. It is found that $p^*_T$ increases monotonously as $\chi_1$ decreases. There is then a threshold temperature $T_0$ below which the equilibrium $p^*_T$ is outside the range (\ref{eq:p_range}), it is marked by: 
\be
p^*_{T_0} = p_+
\ee 
In the limit $\epsilon \ll 1$, the rescaled threshold temperature is given by an order 1 constant to the leading order: 
\be\label{eq:T_bound}
\chi^-_1 = \frac{\pi T_0}{q_1^{5/2}} = \frac{32 \sqrt{58-41 \sqrt{2}}}{5(2 \sqrt{10+\sqrt{2}}+3 \log [1+2 \sqrt{2}-2 \sqrt{2+\sqrt{2}}])} \sim \mathcal{O}(1)
\ee
We interpret this bond as follows. For $\chi_1<\chi^-_1$ the mixed ensembles $\rho^{\beta}_{q_1,q_3}$ does not have a well-defined gravity dual -- at least not described by a one-zone black hole. It is also found that the thermodynamic entropy $S$ decreases monotonously with increasing $p$ in the range (\ref{eq:p_range}). We can therefore deduce that at fixed $(q_1,q_3)$, there is a minimum thermodynamic entropy $S^{\text{min}}= \pi\sqrt{u^{\text{min}}_0}/2G_N$ that a one-zone black hole in the micro-canonical shell can have. It is reached at the threshold temperature $\chi^-_1$. In the limit $\epsilon \ll 1$, the minimum entropy can be computed to the leading order in $\epsilon$ as:
\be
 \sqrt{u_{0}^{\text{min}}} = \sqrt{q_1}\left(1- B\epsilon^2+..\right)
\ee
where the constant coefficient $B$ is given by:
\be
B =\frac{3}{16}\left(3+2 \sqrt{2}\right)\left(-2+\sqrt{2}+\sqrt{2-\sqrt{2}} \tanh^{-1}\left[\sqrt{2-\sqrt{2}}\right]\right) 
\ee
We clarify that $u^{\text{min}}_0$ does not necessarily give the minimum thermodynamic entropy that $\rho^{\beta}_{q_1,q_3}$ can have. For $\chi_1<\chi^-_1$, the mixed ensemble is not described by a one-zone black hole, computing its thermodynamic entropy is therefore beyond the current scope.

\section{Renyi entropies from the gluing construction}\label{sec:renyi}
We now proceed to the computation of holographic Renyi entropies. We are interested in the case of the entangling sub-region $A$ being a large interval on a circle, and the state $\rho$ being an ensemble at fixed KdV charges. For the purpose of being self-contained, we first quickly recall some ingredients of the computation in a more general context. 
\subsection{Review: cosmic-brane backreaction}
The Renyi entropy is defined as: 
\be
S^\psi_n(A) = \frac{1}{1-n}\ln{\text{Tr}\rho^n_A},\;\;\rho_A = \text{Tr}_{\bar{A}}\rho^\psi 
\ee 
Through AdS/CFT, we need to perform a bulk computation of the boundary partition function defined on a branched manifold $\Sigma^n_A$ glued across the sub-region $A$, which specifies the boundary condition for the bulk:
\be
Z_{CFT}(\Sigma^n_A) = \text{Tr} \rho^n_A = Z_{grav}(\Sigma^n_A) 
\ee
To compute $Z_{grav}(\Sigma^n_A)$ one then looks for a particular bulk saddle $\mathcal{B}^n$ such that: 
\be
\partial\mathcal{B}^n = \Sigma^n_A 
\ee
In addition, the asymptotic boundary conditions for the bulk fields are specified by replicating $n$ times those of the state $\psi$, viewed as the bulk dual. The bulk path-integral thus enjoys a $\mathds{Z}_n$ replica symmetry in terms of the boundary conditions, If such a symmetry is inherited by the leading saddle $\mathcal{B}_n$, we can consider its quotient geometry: $\tilde{\mathcal{B}}_n = \mathcal{B}_n/\mathds{Z}_n$. The partition functions are simply related by a factor of $n$: 
\be
Z_{grav}(\mathcal{B}_n)=nZ_{grav}(\tilde{\mathcal{B}}_n)
\ee
The quotient geometry $\tilde{\mathcal{B}}_n$ can be effectively obtained by inserting a co-dimension two defect $\Sigma_n$, i.e. a cosmic-brane, into the bulk state $\psi$ and allow it to backreact \cite{Dong:2016}. The tension $T_n$ of the cosmic brane is related to the Renyi index $n$ via: 
\be
T_n = \frac{n-1}{4n G_N} \;\;\;
\ee      
In the limit $n\to 1$, the cosmic-brane becomes tensionless and simply finds the minimal area configuration in the original geometry, extracting the leading order in $n-1$ effect then gives the RT formula.  

To actually compute the Renyi entropy from the glued solution, it is more convenient to first compute the intermediate quantity called the refined Renyi entropy \cite{Dong:2016}, defined by: 
\be 
\tilde{S}_n(A) = n^2 \partial_n \left(\frac{n-1}{n}S_n(A)\right) = -n^2 \partial_n \left(\frac{1}{n} \ln{\text{Tr}\rho^n_A}\right)
\ee
In holography, this quantity has the advantage of being computed directly by the area of the back-reacted cosmic brane: 
\be
\tilde{S}_n(A) = \frac{\text{Area}(\Sigma^n_A)|_{\mathcal{B}_n} }{4G_N}
\ee
instead of the bulk partition function defined by $\mathcal{B}_n$. From the refined Renyi entropy one can integrate w.r.t $n$ to recover the original Renyi entropy: 
\be\label{eq:int_refined_renyi}
S_n(A) = \frac{n}{n-1}\int^n_1 \frac{\tilde{S}_{\tilde{n}}(A)}{\tilde{n}^2}d\tilde{n} 
\ee

\subsection{High-density limit: the gluing construction}
In general, using the cosmic-brane prescription to actually compute the Renyi entropies is a formidable task -- one needs to solve for the fully backreacted geometry. Further compromise needs to be conceded in order to make progress, e.g. computing the perturbation expansion in small $n-1$ \cite{Dong:2016n1,Bianchi:2016n2} or short distance $\ell$ for the subsystem interval \cite{BinChen:2013,BinChen:2016,Guo:2018,Lin:2016}. The difficult part lies in having to deal with the interplay between cosmic-brane backreaction in the bulk and the asymptotic boundary condition related to the state specification. We are interested in the regime where the KdV charge densities are much larger than the appropriate powers of the inverse subsystem size $L_A$, which is a finite fraction $f$ of the total system size $L$: 
\be
\langle \mathcal{Q}_{2k-1} \rangle_h \sim \left(\frac{h}{c}\right)^{k} L_A^{1-2k},\;\; L_A = f L
\ee
On the other hand, we do not assume anything particular about the Renyi index $n$. We call this the high charge density limit for the KdV charges. The holographic Renyi entropy in the similar limit of the energy micro-canonical ensemble was considered in \cite{Dong:2018}, in which a back-reacted solution $\tilde{B}_n$ was constructed explicitly using a gluing procedure. We shall follow a similar procedure in constructing the back-reacted solution. 

\begin{figure}
    \centering
    \includegraphics[width=0.8\textwidth]{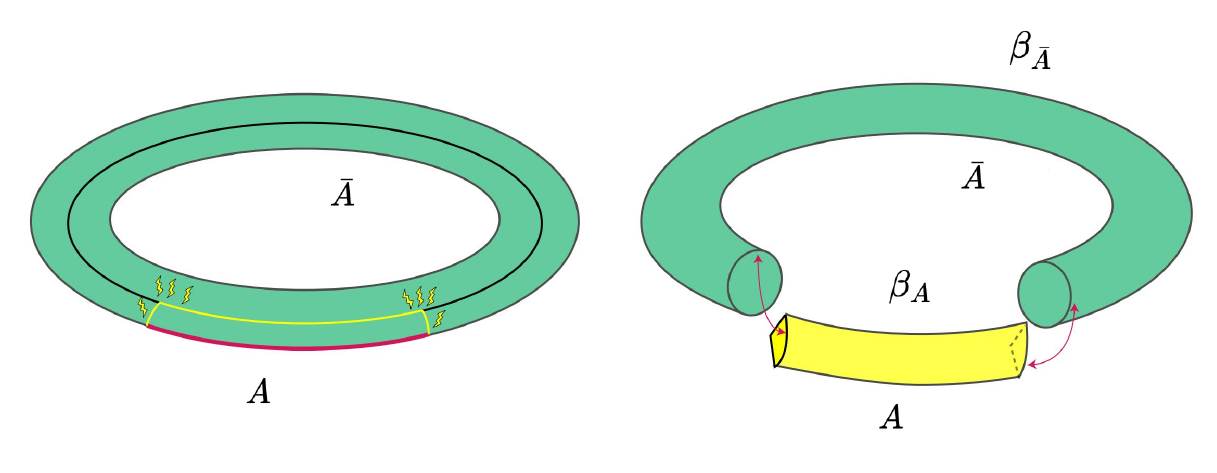}
    \caption{An illustration for glued solution}
    \label{fig:gluedSolution}
\end{figure}

The main idea underlying the gluing construction in \cite{Dong:2018} comes from the following considerations. For simplicity we consider the case of AdS$_3$/CFT$_2$ as in this paper, although the construction in \cite{Dong:2018} works in general dimensions. In the high energy density limit, the Euclidean geometry of the black hole solution fills the asymptotic boundary that is torus whose contractible thermal circle $\beta$ is much smaller than the non-contractable spatial circle of length $L$. Upon inserting a cosmic-string ending on the end points $\partial A$ of a finite interval $L_A = f L$, the back-reaction will equilibrate away from the end points, i.e. producing local geometry well approximated by that of a global black hole solution. If we choose to neglect the details near $\partial A$, the full geometry can be approximated as two segments of black hole solutions, one along $A$ and the other along $\bar{A}$, glued together at the junction $\partial A$ subject to some matching condition, see figure \ref{fig:gluedSolution}.

The matching condition reflects the effect of the cosmic-string insertion, or equivalently the smoothness condition of the bulk saddle $\mathcal{B}_n$ before taking the quotient. By neglecting the details near the junction, only the global constraint that the thermal circle lengths in $\mathcal{B}_n$ must match across $\partial A$ remains, which in the back-reacted quotient geometry $\tilde{\mathcal{B}}_n$ implies the following relation between the black hole temperatures $\beta_A, \beta_{\bar{A}}$  of the two segments: 
\be
\beta_A = n \beta_{\bar{A}} 
\ee
We shall refer to this as the gluing condition. If we are interested in the canonical ensemble at fixed $\beta$, the corresponding glued solution is directly given by: 
\be \label{eq:canonical_match}
\beta_A = n \beta, \;\;\beta_{\bar{A}} = \beta
\ee
On the other hand for the micro-canonical ensemble at fixed total energy $E$, the glued solution is obtained by solving for $\beta$ in (\ref{eq:canonical_match}) via the additional matching condition: 
\be\label{eq:micro_match}
f \langle\mathcal{E}\rangle_{n\beta} + (1-f) \langle \mathcal{E}\rangle_{\beta} = E
\ee
where $\langle \mathcal{E}\rangle_\beta$ is the energy expectation value, i.e. ADM mass of the black hole, at temperature $\beta$. The matching condition (\ref{eq:micro_match}) is basically imposing the asymptotic boundary condition encoding the original state: 
\be
\rho^\psi = \rho^{micro}_E
\ee 
while including the cosmic-string backreaction $\beta_A = n \beta_{\bar{A}}$, simplified in the context of the gluing construction. In fact, the original holographic content has been so minimized that one expects the gluing and matching conditions (\ref{eq:canonical_match},\;\ref{eq:micro_match}) apply to broader contexts featuring similar limits, see \cite{Grover:2017} for the case of chaotic energy eigenstates. In appendix (\ref{sec:gluing_derivation}) we supply additional arguments for (\ref{eq:canonical_match},\;\ref{eq:micro_match}) based on finite dimensional intuitions. The argument there indeed reflects the agreement between the cosmic-brane proposal and the more general diagonal approximation used in the companion paper \cite{KdVETHgeneral}, see also \cite{Dong:2023bfy} for more discussions on this.

More quantitatively, by neglecting the details near the junction boundary $\partial A$, one is essentially focusing only on the volume-dependence of the Renyi entropy, i.e. extracting the contribution that scales like: 
\be\label{eq:volume_law}
S_n(A) \propto L
\ee
We should clarify that by volume-scaling it does not necessarily mean $S_n(A)\propto L_A$ -- there could be prefactor in (\ref{eq:volume_law}) that depends non-linearly on $f$. One can therefore summarize the validity of the gluing construction as follows: in the high charge density limit $E\gg 1/L$, the Renyi entropy of a finite fractional interval $L_A = f L$ is dominated by a contribution that scales with the total volume $L$, and it is this contribution that can be captured by the gluing construction. 

From the $n$-dependent solution $\beta_n$ of (\ref{eq:micro_match}), the volume-scaling part of the refined entropy is simply given then by the partial horizon area from the segment of the black hole solution along $A$.:
\be\label{eq:refined_micro}
\tilde{S}_n(A) = f S_{th}(n\beta_n) 
\ee 
As was pointed out in \cite{Dong:2018}, the integration over $n$ for computing $S_n(A)$ can in fact be done in the following closed form: 
\be\label{eq:renyi_integrate}
S_n(A) =  \frac{f S_{th}(n\beta_n)+(1-f)n S_{th}(\beta_n)-n S_{th}(\beta_1)}{1-n}
\ee
This can be verified by first checking $\lim_{n\to 1}\left[\frac{n-1}{n} S_n(A)\right] =0$ and then computing the following derivative in $n$:
\bea
\partial_n \left[\frac{n-1}{n}S_n(A)\right] = \frac{f s_{th}(n\beta_n)}{n^2} - \left[\frac{f}{n} \left(\partial_\beta S_{th}\right)_{n\beta_n} \partial_n \left(n \beta_n\right) + (1-f) \left(\partial_\beta S_{th}\right)_{\beta_n} \partial_n \beta_n\right]\nonumber
\eea
The terms inside the square bracket cancel by the thermodynamic relation: 
\be
\frac{1}{T}\frac{d \mathcal{E}}{dT} = \frac{dS}{dT} \to \frac{d \mathcal{E}}{d\beta} = \frac{dS}{d\beta} \frac{1}{\beta}
\ee
in conjunction with the matching condition equation (\ref{eq:micro_match}) for $\beta_n$. Therefore the following differential relation holds: 
\be
\partial_n \left[\frac{n-1}{n}S_n(A)\right] = \frac{f s_{th}(n\beta_n)}{n^2}  = \frac{\tilde{S}_n(A)}{n^2}
\ee
in accordance with Eq (\ref{eq:int_refined_renyi}).

It may be worth discussing the range of the Renyi-index $n$ to which the gluing construction is applicable. To this end we perform the following rough estimate. In order for the gluing construction to be a good approximation, $(n-1)$ also needs to be parametrically bounded from below. The effectiveness of the approximation requires the action contribution of the cosmic-brane to parametrically outweigh the corresponding bulk contribution from near the entangling surface $\partial A$-- the latter is neglected in the gluing construction. In very crude terms, this requires that:  
\be\label{eq:validity} 
\left(\frac{n^2-1}{n}\right) f S_{th} \gg \left(\beta \mathcal{E}\right) \zeta 
\ee 
The left hand side represents the cosmic-brane effective action, and the right hand represents the bulk action within a characteristic length scale $\zeta$ near $\partial A$. To be more explicit, we can make the following estimates:
\be
S_{th} \sim \sqrt{\mathcal{E}},\;\; \beta \sim 1/\sqrt{\mathcal{E}},\;\;\zeta \sim \beta
\ee 
Then (\ref{eq:validity}) parametrically corresponds to requiring that: 
\be
n-1 \gg \mathcal{E}^{-1/2}
\ee
This lower bound is therefore invisible to us in the high density limit. The nature of this bound is conceptually similar to the requirement of $(n-1) \gg 1/c$ implicitly assumed in the cosmic-brane prescription -- such that the classical action contribution from the cosmic-brane parametrically outweighs the quantum corrections that is neglected. It is worth mentioning that despite this subtlety, the limits of $n\to 1$ and $c\to \infty$ are usually assumed to be commuting in most holographic contexts. However, there are scenarios \cite{Akers:2020pmf,Dong:2023xxe} where they do not commute and the order of limits is indeed important.    

\subsection{Gluing construction at fixed KdV charges}\label{sec:renyi_KdV}
Now we generalize the gluing construction to the context of ensembles at a finite number of fixed KdV charges. The goal is to compute the Renyi entropy in ensembles at fixed KdV charges $\langle\mathcal{Q}_{2k-1}\rangle =q_{2k-1},\;k=1,...,m$, which we collective denote as $\lbrace q \rbrace$: 
\be
S^{\lbrace q\rbrace}_n(A) = \frac{1}{1-n}\ln{\text{Tr}\rho^n_A(\lbrace q\rbrace)},\;\;\;\rho_A(\lbrace q \rbrace) = \text{Tr}_{\bar{A}} \rho^{micro}_{\lbrace q\rbrace}
\ee 
along a finite interval $L_A = fL$. We begin with the micro-canonical ensemble at these KdV charges. By similar lines of argument, the gluing construction is an effective approximation in the large charge-density limit: 
\be
q_{2k-1} \gg L^{1-2k},\;\;k=1,...,m
\ee
Given that $q_{2k-1}\geq q_1^{k}$, this would follow if only the high energy density limit is fulfilled: 
\be
E = \langle q_1\rangle \gg L^{-1} 
\ee
Again we are only focusing the $L$-scaling part of the result, i.e. ignoring contributions coming from the junction effects near $\partial A$. 

More specifically, we propose to construct the back-reacted geometry $\mathcal{B}_n$ computing $S^{\lbrace q\rbrace}_n(A)$ as follows. We glue two segments of black hole geometries long $A$ and $\bar{A}$ respectively. The segments are characterized by two sets of KdV chemical potentials $\lbrace \mu_A\rbrace$  and $\lbrace \mu_{\bar{A}}\rbrace$ -- locally they are the black hole solutions describing the GGEs $e^{-\sum_k \tilde{\mu}_A^{2k-1} \hat{Q}_{2k-1}}$ and $e^{-\sum_k \tilde{\mu}_{\bar{A}}^{2k-1} \hat{Q}_{2k-1}}$ respectively. The natural gluing conditions to be imposed at the junction are: 
\be\label{eq:KdV_glue}
\mu^{2k-1}_A = n \mu^{2k-1}_{\bar{A}} = n \mu^{2k-1},\;\;k=1,...,m
\ee
while the asymptotic boundary conditions characterizing $\rho^{micro}_{\lbrace q\rbrace}$ now impose additional matching conditions for each of the fixed KdV charges $q_{2k-1}$: 
\be\label{eq:KdV_match}
f \langle \mathcal{Q}_{2k-1}\rangle_{n\lbrace\mu\rbrace} + (1-f) \langle \mathcal{Q}_{2k-1}\rangle_{\lbrace\mu\rbrace} = q_{2k-1},\;\;k=1,...,m
\ee
where $\langle \mathcal{Q}_{2k-1}\rangle_{\lbrace \mu\rbrace}$ is the $k$-th KdV charge density evaluated in the black solution describing $e^{-\sum_k \tilde{\mu}_A^{2k-1} \hat{Q}_{2k-1}}$. From these we should solve for $(\lbrace \mu_n\rbrace)$, the refined and ordinary Renyi entropies are given analogously to (\ref{eq:refined_micro},\;\ref{eq:renyi_integrate}): 
\bea\label{eq:renyi_integrate_KdV}
\tilde{S}^{\lbrace q\rbrace}_n(A) &=& f S_{th}(n\lbrace\mu_n\rbrace)  \nonumber\\
S^{\lbrace q\rbrace}_n(A) &=& \frac{f S_{th}(n\lbrace\mu_n\rbrace)+(1-f)n S_{th}(\lbrace\mu_n\rbrace)-n S_{th}(\lbrace\mu_1\rbrace)}{1-n}
\eea
The integral over $n$ is done similarly by invoking the extended thermodynamic relations together with the matching conditions (\ref{eq:KdV_match}):
\be
\sum^{m}_{k=1}\mu^{2k-1}\frac{d\langle \mathcal{Q}_{2k-1}\rangle}{d\mu^{2j-1}} = \frac{dS}{d\mu^{2j-1}},\;\;\;j=1,...,m
\ee 
So far the analysis has been a straightforward generalization of the computation for the ordinary micro-canonical ensemble in energy. Before we end the general discussion and turn to more concrete cases, let us clarify some subtleties that arises due to the nature of the KdV-charged black hole solutions. 

Firstly, for GGEs with $k$ non-zero KdV chemical potentials, as discussed before the generic finite-zone black holes are labeled by $k-1$ free parameters. When carrying out the gluing construction, a glued solution is physical only when the black hole geometries along each of the segments $A$ and $\bar{A}$ are the dominant Euclidean saddle-points in the corresponding GGEs at $(n\beta,\mu_1,...,\mu_k)$ and $(\beta,\mu_1,...,\mu_k)$ respectively. Otherwise their charges do not represent the correct expectation values in the GGEs when satisfying matching conditions. However, it is beyond the scope of this work to fully identify the dominant Euclidean saddle-point systematically. For our purpose, we will for the most part confine the analysis to include only BTZ and one-zone black holes. We will refrain from including black hole solutions with more than two zones. Roughly speaking, black holes with a larger number of zones excite more oscillatory modes and thus tend to have higher energies, they are therefore less likely to be thermodynamically stable. These are intuitions subject to closer scrutiny, we leave them for future investigations.

Secondly, there may be cases where multiple glued solutions exist that are all physical in the sense just described, i.e. each consisting of segments from the dominant Euclidean saddle of the corresponding GGEs along $A$ and $\bar{A}$ respectively. They should then all be considered as quotients of legitimate Euclidean saddle-points for the partition function on the replica manifold that computes the Renyi entropy: 
\be
Z(\Sigma^n_A) = \text{Tr} \rho_A^n
\ee  
The glued solution to be identified as the dominant Euclidean saddle-point of $Z(\Sigma^n_A)$ should then be determined by minimizing the corresponding free energy, which in this case is proportional to the Renyi entropy $S_n(A)$. 
\begin{figure}[ht]
    \centering
    \begin{minipage}[h]{0.45\linewidth}
        \includegraphics[width=\linewidth]{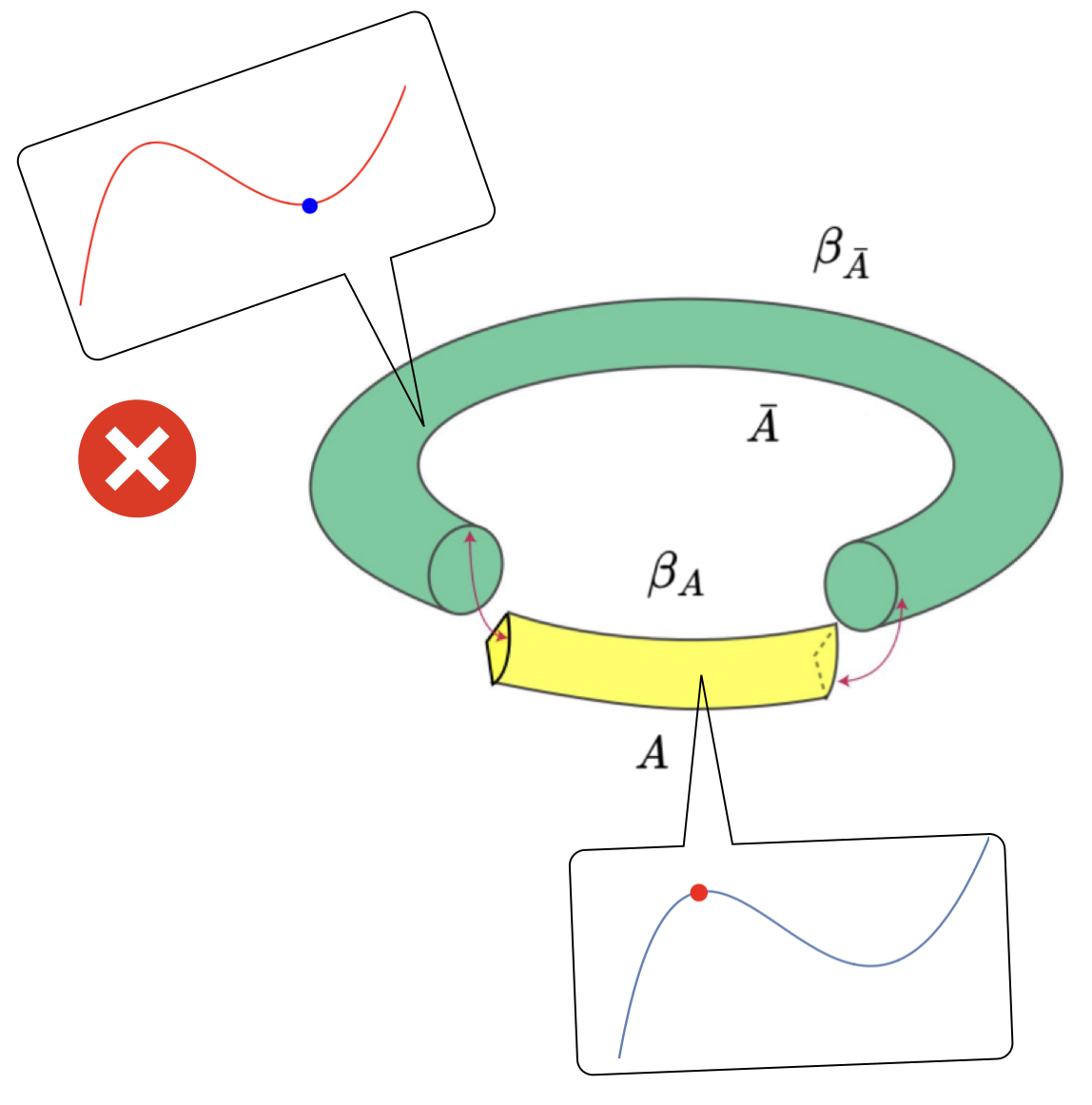}
        \label{}
    \end{minipage}
    \begin{minipage}[h]{0.45\linewidth}
         \vspace{-0.3cm}
        \includegraphics[width=\linewidth]{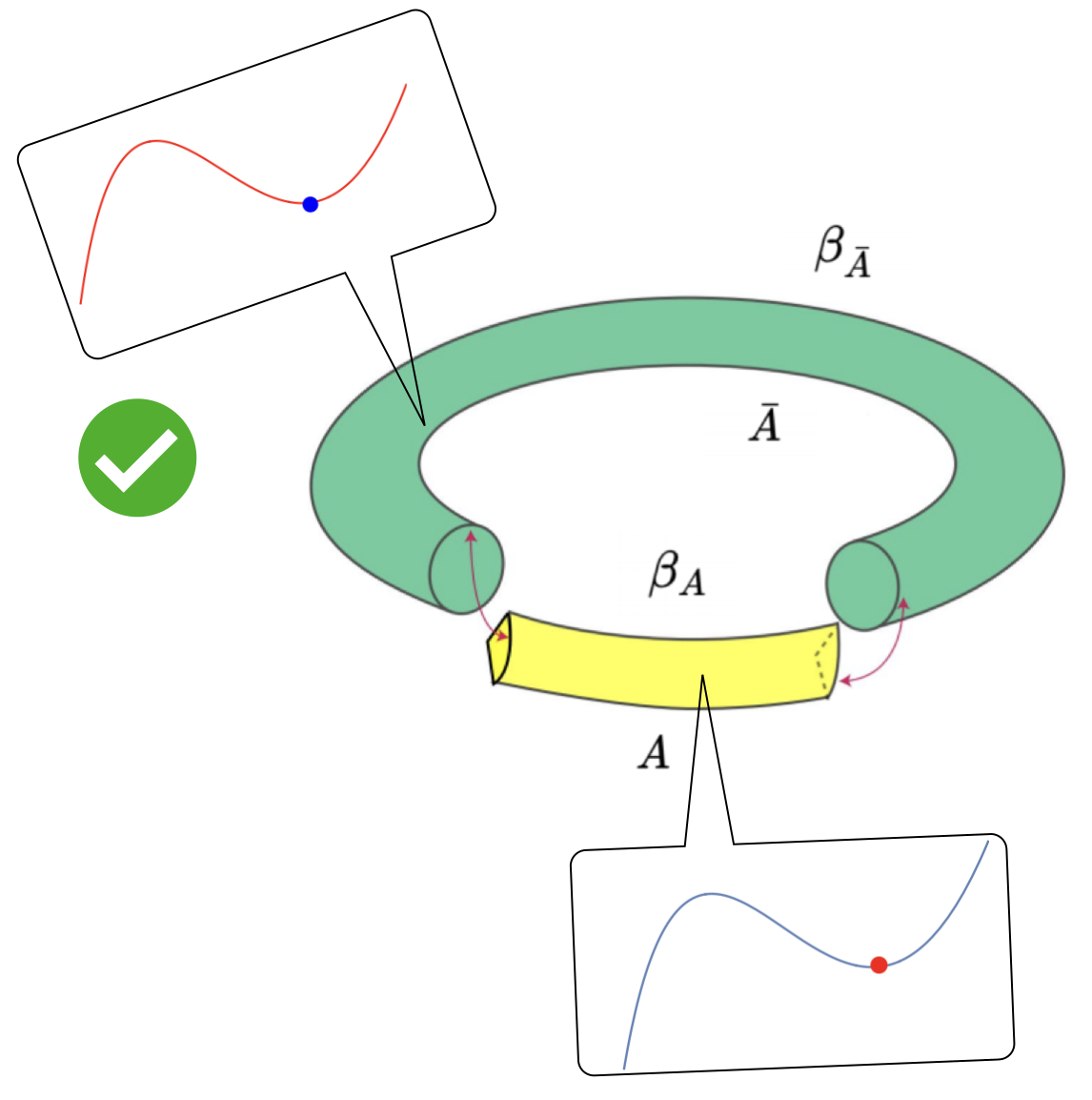}
        \label{}
    \end{minipage}
    \caption{In a legitimate glued solution, each consisting of segments should come from the dominant Euclidean saddle of the corresponding GGEs along $A$ and $\bar{A}$ respectively.}
\end{figure}

Thirdly, as has been suggested in (\ref{eq:KdV_glue}) by the index range, the glued solutions should be constructed using black holes segments from GGEs with only the first $m$ KdV chemical potentials $\lbrace\mu^{2k-1},k=1,...,m\rbrace$ turned on. This is to be compatible with the underlying micro-canonical ensemble fixing the first $m$ total KdV charges: 
\be
\rho^{micro}_{\lbrace q\rbrace} = \mathcal{N}^{-1} \hat{P}_{\lbrace q\rbrace} 
\ee
However, there may be cases where under such constraints it is impossible to find glued solutions that are valid saddle-points of the Renyi entropy computations in the sense just discussed. We can then choose to consider more general GGEs, i.e. those of the form with $m'>m$ KdV chemical potentials turned on $\lbrace \mu^{2k-1},\;k=1,...,m'\rbrace$. However, we shall interpret the additional chemical potentials $\lbrace \mu'\rbrace=\lbrace \mu^{2k-1},m<k\leq m'\rbrace$ as physical parameters, i.e. not to be solved from gluing/matching conditions. Accordingly the glued solutions constructed from these GGEs should be interpreted as saddle-points responsible for computing the Renyi entropies in the following ensembles: 
\be \label{eq:mixed_integral_renyi}
\rho^{\lbrace \mu'\rbrace}_{\lbrace q\rbrace} = \mathcal{N}^{-1}\hat{P}_{\lbrace q\rbrace} e^{-\sum_{k>m} \tilde{\mu}^{2k-1} \hat{Q}_{2k-1}}
\ee
Namely, in these ensembles the states within the micro-canonical shell labeled by $\lbrace q\rbrace$ do not contribute with equal weights as the micro-canonical ensemble does, but instead are weighted according to their higher KdV charges $\lbrace\mathcal{Q}_{2k-1},\;m<k\leq m'\rbrace$. These are the generalized form of the mixed ensemble considered in section (\ref{sec:mmGGE}). Related to this, the Renyi entropy formula (\ref{eq:renyi_integrate_KdV}) should be modified accordingly by replacing the thermodynamic quantities used: 
\be\label{eq:renyi_integrated_mm}
S_{th}\to S_{th}-\sum_{k>m}\tilde{\mu}^{2k-1} \hat{Q}_{2k-1}
\ee 

Lastly, generic finite-zone solutions are inhomogeneous, i.e. the classical stress energy field $u(\varphi)$ varies along the spatial circle. Such inhomogeneity may therefore add additional subtleties to the gluing construction, e.g. the location of the junction $\partial A$ inside each black hole segment may become relevant. Although most of the explicit gluing computations in later sections only concern the BTZ geometries, let us make some comments regarding this issue. We argue that in the limit of our interest, we can neglect such inhomogeneity and treat the KdV charge densities as homogeneously distributed along the spatial circle, i.e. the partial KdV charges along the segments are simply given by $f \langle \mathcal{Q}_{2k-1}\rangle$ and $(1-f)\langle \mathcal{Q}_{2k-1}\rangle$. There are two reasons for such an approximation. Firstly, in the high charge density limit a generic solution has its typical scale of density oscillation much smaller than the subsystem size. Therefore the details of density distribution reflecting the inhomogeneity is subleading to the limit we are interested. At a more fundamental level, in \cite{Dymarsky:2020} it was argued that the gravity dual of the CFT GGE, even at fixed zone parameters, does not correspond to an individual finite-zone solution, instead one should statistically average over the Jacobian manifold of the phase space related to the finite-zone solution. This includes in particular the images under translation. Upon averaging, the details related to the inhomogeneity are obliviated. 

\section{Holographic Renyi entropy at fixed $\langle\mathcal{Q}_1\rangle$ and $\langle \mathcal{Q}_3\rangle$}\label{sec:main}
In this section, we apply the prescription in (\ref{sec:renyi_KdV}) to a concrete setting. We study in detail the ensembles fixing only the first two KdV charges: 
\be
\langle \mathcal{Q}_1\rangle = q_1,\;\;\langle \mathcal{Q}_3\rangle = q_3 
\ee 
In the high density limit we impose that $q_1\gg 1,\; q_3 \gg 1$. A glued solution consists of two segments of black hole solutions from a GGE with chemical potentials collectively denotes as $\mu = \lbrace \mu\rbrace$. The gluing/matching conditions combined take the form: 
\bea\label{eq:case_match}
&&f\langle \mathcal{Q}_1 \rangle_{n\beta,\mu} + (1-f)\langle \mathcal{Q}_1 \rangle_{\beta,\mu} = q_1\nonumber\\
&&f\langle \mathcal{Q}_3 \rangle_{n\beta,\mu} + (1-f)\langle \mathcal{Q}_3 \rangle_{\beta,\mu} = q_3
\eea  
We will consider the micro-canonical ensemble: 
\be
\rho^{micro}_{q_1,q_3} = \mathcal{N}^{-1}\hat{P}_{q_1,q_3} 
\ee
and will later extend to the mixed ensembles: 
\be 
\rho^{\beta}_{q_1,q_3} =\mathcal{N}^{-1} \hat{P}_{q_1,q_3}\;e^{-\beta \hat{Q}_5} 
\ee

\subsection{Glued BTZ geometries}\label{sec:glued_BTZ}
We begin with the computation in $\rho^{micro}_{q_1,q_3}$. The corresponding Renyi entropy $S^{q_1,q_3}_n(A)$ is then computed by constructing the glued solutions using GGEs of the form: 
\be\label{eq:GGE_q3}
\rho_{GGE}(\beta,\mu) = \mathcal{N}^{-1} e^{-\beta \left(\hat{Q}_3+\tilde{\mu} \hat{Q}_1\right)} 
\ee
As has been discussed in section (\ref{sec:one-zone}), these GGEs only have the BTZ black holes as stable Euclidean saddle-points, the one-zone black holes have negative temperatures. We therefore compute the glued-solutions using only BTZ black holes. When the black hole segments are both BTZs, their KdV charge densities satisfy: 
\be
\langle \mathcal{Q}_3\rangle_{n\beta,\mu} = \langle \mathcal{Q}_1\rangle^2_{n\beta,\mu} = q_A^2,\;\;\langle \mathcal{Q}_3\rangle_{\beta,\mu} = \langle \mathcal{Q}_1\rangle^2_{\beta,\mu} = q_{\bar{A}}^2 
\ee
In this case, $(q_A,\;q_{\bar{A}})$ can be solved from the matching conditions alone:
\be 
f q_A + (1-f) q_{\bar{A}} = q_1,\;\;f q_A^2+ (1-f) q_{\bar{A}}^2 = q_3 
\ee 
They are explicitly given by:
\be \label{eq:2_branches}
q^{\pm}_A= q_1 \mp \left(1/f-1\right)^{1/2}\sqrt{\Delta q_3},\;\; q^{\pm}_{\bar{A}} = q_1 \pm \left(1/f-1\right)^{-1/2}\sqrt{\Delta q_3}
\ee
where we have defined: 
\be
\Delta q_3 = q_3 - q_1^2 > 0 
\ee
We are interested in the effect of $\Delta q_3$ that are visible in the high density limit, so we always assume that $\Delta q_3 \propto q_1^2\gg 1$. There are two branches of glued BTZ solutions (\ref{eq:2_branches}): the $(+)$ branch exists for $\Delta q_3/q_1^2<(1/f-1)^{-1}$; the $(-)$ branch exists for $\Delta q_3/q_1^2<1/f-1$. For each branch the KdV charges are $n$-independent. The $n$-dependence comes from the gluing conditions, and in this case they only determine the GGE parameters describing the BTZ geometries. They are fixed by requiring that:
\be 
q_A = \langle \mathcal{Q}_1\rangle^{BTZ}_{n\beta_n,\mu_n},\;\;q_{\bar{A}} = \langle \mathcal{Q}_1\rangle^{BTZ}_{\beta_n,\mu_n}
\ee
For BTZ geometries in (\ref{eq:GGE_q3}) they become the following algebraic equations: 
\be 
\frac{2\pi}{\sqrt{q_A}} = n\beta_n\left(2q_A+\mu_n\right),\;\; \frac{2\pi}{\sqrt{q_B}} = \beta_n\left(2q_B+\mu_n\right)
\ee
The solutions for $(\beta,\mu)$ are given simply by:
\be \label{eq:glued_BTZ_GGE}
\beta_n = \frac{\pi\left(n \sqrt{q_A}-\sqrt{q_{\bar{A}}}\right)}{n \sqrt{q_A q_{\bar{A}}}\left(q_{\bar{A}}-q_A\right)},\;\;\mu_n = \frac{2n q_A^{3/2}-2q^{3/2}_{\bar{A}}}{\sqrt{q_{\bar{A}}}-n\sqrt{q_A}}
\ee 
One can check that of the two branches in (\ref{eq:2_branches}), only the $q^+_{A,\bar{A}}$ branch: 
\be \label{eq:glued_BTZ_charges}
q_A= q_1 - \left(1/f-1\right)^{1/2}\sqrt{\Delta q_3},\;\; q_{\bar{A}} = q_1 + \left(1/f-1\right)^{-1/2}\sqrt{\Delta q_3}
\ee
corresponds to BTZ segments with positive temperatures $\beta>0$. From these $\mathcal{Q}_1$ charges, one can directly obtain the Renyi entropies:
\bea\label{eq:KdV_renyi}
\tilde{S}_n(A) &=& \frac{f \pi \sqrt{q_A }}{2 G_N}\nonumber\\
S_n(A) &=& \frac{\pi n \sqrt{q_A}}{2G_N(n-1)}\left[\sqrt{\frac{u_0}{q_A}}-(1-f)\sqrt{\frac{q_{\bar{A}}}{q_A}}-\frac{f}{n}\right]
\eea
where $u_0$ is the thermal entropy of the original micro-canonical ensemble $\rho^{micro}_{q_1,q_3}$. An immediate problem with the glued BTZ solution (\ref{eq:glued_BTZ_charges},\;\ref{eq:glued_BTZ_GGE}) is that the BTZ segments have positive temperatures only for $n\geq n_c$:
\be\label{eq:micro_ncrit}
n_c=\sqrt{q_{\bar{A}}/q_A}
\ee 
For $n<n_c$, the BTZ segments are of negative temperatures, and thus (\ref{eq:glued_BTZ_charges},\;\ref{eq:glued_BTZ_GGE}) ceases to be a valid saddle-point geometry of the cosmic-brane back-reaction. On the other hand, we know that for GGEs of the form (\ref{eq:GGE_q3}) the only stable Euclidean saddle-point are the BTZ black holes. As a result, for $n<n_c$ we run out of ingredients to construct a glued solution using well-defined black hole segments -- the cosmic-brane back-reaction on $\rho^{micro}_{q_1,q_2}$ no longer yields a well-defined bulk dual. We perceive this as a consequence of the fact discussed in section (\ref{sec:mmGGE}) that $\rho^{micro}_{q_1,q_3}$ itself does not have well-defined bulk dual, which is revealed in the limit of diminishing back-reaction $n\to 1$. We remark that a phase transition at a critical Renyi index $n_c$ precisely equal to (\ref{eq:micro_ncrit}) was also discovered in the companion paper \cite{KdVETHgeneral} using more general approaches, but concerning a different class of ensembles. 

As commented at the end of section (\ref{sec:renyi_KdV}), for lower $n<n_c$ we could remedy the situation by considering glued solutions in more general ensembles. By doing this, the back-reacted solution is likely to have a well-defined gravity dual. The simplest choice is to add an additional chemical potential for $\hat{Q}_5$, which we view as the new temperature, and solve the gluing/matching equation by black holes in the GGEs: 
\be\label{eq:GGE_q5}
\rho = \mathcal{N}^{-1}\;e^{-\beta \mathcal{H}},\;\;\mathcal{H}=\hat{Q}_5+\mu^3 \hat{Q}_3+\mu^1 \hat{Q}_1
\ee
In section (\ref{sec:thermo_onezone}) we have studied in details the properties of the BTZ black holes in these GGEs. As a result of making this modification, we are now effectively computing the Renyi entropies in the mixed ensemble of the form studied in section (\ref{sec:mmGGE}): 
\be\label{eq:mmGGE}
\rho^{\beta}_{q_1,q_3} =\mathcal{N}^{-1} \hat{P}_{q_1,q_3}\;e^{-\beta \hat{Q}_5} 
\ee
The glued BTZ solutions still consist of two branches of charge densities (\ref{eq:2_branches}) -- they come only from the matching conditions. The GGE parameters are solved by the gluing conditions in terms of the BTZ saddle-point equations in (\ref{eq:GGE_q5}): 
\be
\frac{2\pi}{\sqrt{q_A}} = n\beta \left(3 q_A^2 + 2\mu^3_n q_A +\mu^1_n\right),\;\; \frac{2\pi}{\sqrt{q_{\bar{A}}}} = \beta \left(3 q_{\bar{A}}^2 + 2\mu^3_n q_{\bar{A}} +\mu^1_n\right)
\ee
The temperature $T=1/\beta>0$ is fixed as the physical parameter. In this case, both branches of (\ref{eq:2_branches}) consists of BTZ segments with positive temperatures. We begin with the $(+)$ branch (\ref{eq:glued_BTZ_charges}), and will discuss the other one subsequently. The remaining GGE parameters $(\mu^3_n,\mu^1_n)$ can then be solved and are given by: 
\bea\label{eq:glued_BTZ_GGE_q5}
\mu^3_n &=& \frac{2\pi\left(\sqrt{q_{\bar{A}}}-n \sqrt{q_A}\right)+3n\sqrt{q_A q_{\bar{A}}}\left(q_{\bar{A}}^2-q_A^2\right)\beta}{2n \sqrt{q_A q_{\bar{A}}} \left(q_A-q_{\bar{A}}\right)\beta}\nonumber\\
\mu^1_n &=& \frac{2\pi\left(n q^{3/2}_A- q^{3/2}_{\bar{A}}\right)+3n\left(q_A q_{\bar{A}}\right)^{3/2}\left(q_A-q_{\bar{A}}\right)\beta}{n \sqrt{q_A q_{\bar{A}}} \left(q_A-q_{\bar{A}}\right)\beta}
\eea
Therefore at fixed $\beta>0$, the glued BTZ solutions (\ref{eq:glued_BTZ_charges},\;\ref{eq:glued_BTZ_GGE_q5}) in the mixed ensemble (\ref{eq:mmGGE}) can stand as smooth bulk solutions to the gluing construction at any $n$. The Renyi entropies from these glued solutions are given analogously by: 
\bea\label{eq:mixed_renyi} 
\tilde{S}_n(A) &=& \frac{f \pi \sqrt{q_A }}{2 G_N}\nonumber\\
S_n(A) &=& \frac{\pi n \sqrt{q_A}}{2G_N(n-1)}\left[\sqrt{\frac{u_0}{q_A}}-(1-f)\sqrt{\frac{q_{\bar{A}}}{q_A}}-\frac{f}{n}\right]\nonumber\\
&-& \frac{\beta n}{4G_N(n-1)}\left[\mathcal{Q}_5 -(1-f)q_{\bar{A}}^3-f q_A^3\right]
\eea
where we have invoked the substitution (\ref{eq:mixed_integral_renyi}) for computing the Renyi entropy in the mixed ensembles, and $(u_0,\mathcal{Q}_5)$ are the entropy and $\mathcal{Q}_5$ expectation value in the original ensemble $\rho^{\beta}_{q_1,q_3}$. The next task is to investigate whether they correspond to the dominant Euclidean saddle of $\text{Tr}\rho_A^n$, in which case (\ref{eq:mixed_renyi}) gives the correct Renyi entropies. The answer depends on the physical parameters, which in total include $(q_1,q_3,n,\beta)$. We will focus on a particular regime for $q_1$ and $q_3$ that we call the \textit{near-primary} regime, to be introduced as follows. 
 
\subsection{Near-primary regime}
As discussed in the introduction, our goal is to study ensembles that resemble the primary states. In terms of the fixed KdV charges $(q_1,q_3)$, we are therefore interested in the cases where they approach to saturate the relation: 
\be
q_3 \to q_1^2
\ee 
We emphasize that in doing this, we keep $\Delta q_3 \propto q_1^2 \gg1 $ visible in the high density limit, it is the ratio:
\be
\epsilon = \sqrt{\Delta q_3}/q_1
\ee 
that we are sending to small values. Notice that if we send $\Delta q_3$ itself to small values, but remain in the classical description, i.e. at the leading order in $c\to \infty$, it describes the BTZ black holes. If we further enforce $\hat{Q}_3 = \hat{Q}_1^2$ exactly on the quantum KdV charges, it describes primary states in the boundary CFTs. For this reason, we will take liberty to call the regime $\epsilon\ll 1$ near-primary, even though $\Delta q_3 \gg 1$. 

From now on let us work in the near-primary regime and focus on the mixed ensemble $\rho^{\beta}_{q_1,q_3}$. We are interested in the range of $n$ in which the glued BTZ solutions (\ref{eq:glued_BTZ_charges},\;\ref{eq:glued_BTZ_GGE_q5}) provide the dominant saddle-point for the computation of the Renyi entropy. The Renyi entropies are then given by (\ref{eq:mixed_renyi}). We remind that for the micro-canonical ensemble the range is simply given by:
\be
n>n_c = \sqrt{\frac{q_{\bar{A}}}{q_A}} = 1+ \frac{\epsilon}{2\sqrt{f(1-f)}} +...
\ee 
in the near-primary regime. As discussed at the end of section (\ref{sec:renyi_KdV}), an affirmative answer favoring the glued BTZ can be decomposed into two aspects: (i) it has to be a valid saddle-point in the sense discussed previously; (ii) when multiple saddle-points exist, it has to minimize the Renyi entropy against other possibilities.

\subsubsection{Instability towards $n\to 1$}\label{sec:instability}
Recall that a glued solution is valid if both of the black hole geometries along $A$ and $\bar{A}$ are stable in the corresponding GGEs. For the glued BTZ solution  (\ref{eq:glued_BTZ_charges}, \ref{eq:glued_BTZ_GGE_q5}), this amounts to requiring that both BTZ segments along $A$ and $\bar{A}$ are thermodynamically stable in the GGE with parameters $(n\beta,\mu^3_n,\mu^1_n)$ and $(\beta,\mu^3_n,\mu^1_n)$ respectively. The answer to this question depends on the Renyi index $n$ through the GGE parameters. In what follows we analyze this question as $n$ is varied. 

Let us first think in general about the $n\to 1$ limit of the glued BTZ solutions. It is clear that they cannot persist as good approximations to the back-reacted geometry. This is because that when the cosmic-brane tension vanishes as $n\to 1$, the bulk geometry of the original mixed ensemble should be recovered. It is studied in section (\ref{sec:mmGGE}), and among other properties it is homogeneous with respect to subregions. Therefore the distinction between the $A$ and $\bar{A}$ segments should diminish as $n\to 1$. It is clear that the glued BTZ solution fails to exhibit this, e.g. the charge density difference between $A$ and $\bar{A}$ remains fixed as one takes the limit $n\to 1$: 
\be
q_{\bar{A}} - q_A = \sqrt{\frac{\Delta q_3}{f(1-f)}}
\ee
Related to this, it fails the expectation that in the $n\to 1$ limit $\tilde{S}_n(A)$ should coincide with the von-Neumann entropy $S_{vn}(A)$ of the original ensemble. In our case $S_{vn}(A)$ is simply given by the fractional thermodynamical entropy computed in section (\ref{sec:mmGGE}):
\be
\lim_{n\to 1} \tilde{S}_n(A) = \frac{f \pi \sqrt{q_A}}{2G_N} \neq S_{vn}(A) = \frac{f \pi \sqrt{u_0}}{2G_N}
\ee
Because of this, sufficiently close to $n=1$ the glued BTZ solution has to become unphysical and give ways to other forms of solutions, so as to be consistent with the above considerations. In other words, the BTZs along either of the segments $A$ and $\bar{A}$ has to become unstable in the corresponding GGEs.

We consider two types of instabilities. Firstly, it could become unstable due to additional BTZ solutions in the same GGE but has lower free energies, i.e. unstable via first order phase transitions, we will call these the first order instabilities. Secondly, the BTZ segment could become perturbatively unstable in free energies against nearby one-zone black holes, we will call these the second order instabilities for reasons to be discussed later.  Both have been discussed in (\ref{sec:thermo_onezone}). It is worth pointing out that the first-order instability is guaranteed to be present at $n=1$, where the BTZ segments along $A$ and $\bar{A}$ belong to the same GGE $(\beta,\mu^3_1,\mu^1_1)$. As a result, the charge densities $(h,\bar{h})=(q_{A},q_{\bar{A}})$ correspond to two distinct roots of the same saddle-point equation from extremizing $\mathcal{F}_{BTZ}(h)$:
\be
\frac{2\pi}{\sqrt{h}} = \beta \left(3h^2+2\mu^3_1 h+\mu^1_1\right) 
\ee
It is impossible for both $(h,\bar{h})$ to be the global minimum of the GGE. One of the them has to have higher free energy -- either as a local maximum or as a meta-stable local minimum. This provides a ``backup" channel of instability that prevents the glued BTZ solution from reaching all the way to $n=1$, as expected. 

Now we investigate in details the onset of the instabilities considered. The thermodynamic properties of the BTZ black hole segments along $A$ and $\bar{A}$ depend on the corresponding parameters $(\chi^{\bar{A}}_{1,2},\;\chi^A_{1,2})$ defined in section (\ref{sec:thermo_onezone}). According to (\ref{eq:glued_BTZ_GGE_q5}) they are given by:
\bea\label{eq:glued_BTZ_chis}
\chi^A_1 &=& \left(\frac{\pi T}{n q_A^{5/2}}\right),\;\;\chi^A_2 = \left(\frac{\mu^3_n}{q_A}\right)=\frac{2\chi^A_1 (n_c-n)+3n_c(n_c^4-1)}{2n_c(1-n_c^2)}\nonumber\\
\chi^{\bar{A}}_1 &=& \left(\frac{\pi T}{q_{\bar{A}}^{5/2}}\right),\;\;\chi^{\bar{A}}_2 = \left(\frac{\mu^3_n}{q_{\bar{A}}}\right) = \frac{2\chi^{\bar{A}}_1 (n_c-n)+3n\left(1-n_c^{-4}\right)}{2n\left(n_c^{-2}-1\right)}
\eea

Let us recall some relevant discussions from the stability analysis in section (\ref{sec:thermo_onezone}). For our purpose, a BTZ segment, say along $A$ with $\mathcal{Q}_1$ charge density $h$, is perturbatively unstable against nearby one-zone black holes in the corresponding GGE if both its branches $(h,w^{\pm})$ are deformable and exhibit negative free energy cost $\delta F<0$, as computed in (\ref{eq:dF_sign}). When the BTZ is deformable as the $p\to 1$ limit of one-zone black holes, we have concluded that the $w^+$ branch is always thermodynamically stable. In this case we can pick the $w^+$ branch for the glued BTZ solution and avoid potential instabilities. Therefore the second order instabilities can only happen when the BTZ is deformable as the $p\to 0$ limit of one-zone black holes. This corresponds to when: 
\be\label{eq:pert_inst}
\chi_1>4,\;\;\; \text{max}\lbrace y_{\pm} \rbrace <\chi_2<\tilde{\zeta}_4 = -2\sqrt{\chi_1}-1<0
\ee
where $y_{\pm}$ are the threshold values determined by the roots of (\ref{eq:thresholds}). 

If any of the BTZ segments (\ref{eq:glued_BTZ_chis}), say that along $A$ with $\mathcal{Q}_1$ charge density $h$, comes from a GGE satisfying (\ref{eq:3_BTZ}), it becomes susceptible to the first order instability via bubble nucleation into another BTZ saddle-point with $\mathcal{Q}_1$ charge density $h'$ in the GGE that has lower free energy: 
\bea\label{eq:nonpert_inst}
\mathcal{F}_{BTZ}(h)>\mathcal{F}_{BTZ}(h')
\eea 

At generic values of $n$, multiple types of instabilities in either BTZ segment may coexist. We are interested in the earliest onset of instability starting from sufficiently large $n$, i.e. the maximum value of $n_{cut}$ at which some instability occurs in either BTZ segment (\ref{eq:glued_BTZ_chis}). This value is very important because it provides a cut-off for $n$ below which we can no longer trust \eqref{eq:KdV_renyi}. As we will discuss later, it then reveals important entanglement properties underlying the ensemble $\rho^{\beta}_{q_1,q_3}$ across $A$. 
  
As $n$ is varied, $\chi_1 \equiv \chi^{\bar{A}}_1$ is fixed, and we treat it as representing the temperature $T$. In the near-primary limit $n_c-1\sim \epsilon \ll 1$, we can summarize the results regarding $n_{cut}$ as follows. The details of the analysis can be referred to in the appendix (\ref{app:instabilities}).
\begin{itemize}
\item In the high temperature limit, $n_{cut}$ is dictated by the second order instability along the $\bar{A}$ segment and admits the following expansion in $\eta = 1/\chi_1^{1/4}$: 
\be
n_{cut}\left(\chi_1\right) = n_c -\frac{2(n_c^2-1)}{n_c} \eta^2 + \frac{9 n_c^4-14 n_c^2+5}{2n_c^3} \eta^4+...
\ee
\item In the low temperature regime, $n_{cut}$ is dictated by the first order instability along the $A$ segment, and there is a lower limit $\Delta \chi_1$ at which $n_{cut}$ diverges:
\bea 
n_{cut}\left(\chi_1\right)= \frac{17n_c^4-14 n_c^2+3-6n_c^6}{2n_c^3\left(\chi_1-\Delta \chi_1\right)}+...,\;\;\;\Delta\chi_1=\frac{(n_c^2-1)(3n_c^2-1)}{2n^4}
\eea
For lower temperatures $\chi_1\leq \Delta \chi_1$, the glued BTZ solution becomes invalid for all $n\geq 1$. 
\item There is an intermediate temperature $\chi_c$, at which the first order instabilities are absent along both segments for all $n\geq 1$, and $n_{cut}$ is given by the second order instability along $A$: 
\bea 
n_{cut}(\chi_c)=  1+ \frac{5}{16}\left(n_c-1\right)^3+...,\;\;\chi_c = \frac{1}{2}n_c^{-4}(1+n_c)^3
\eea
This marks the closest to $n=1$ that the lower cut-off $n_{cut}$ can get at fixed $(q_1,q_3)$.
\end{itemize}

\begin{figure}[ht]
	\centering
	\includegraphics[width=0.85\linewidth]{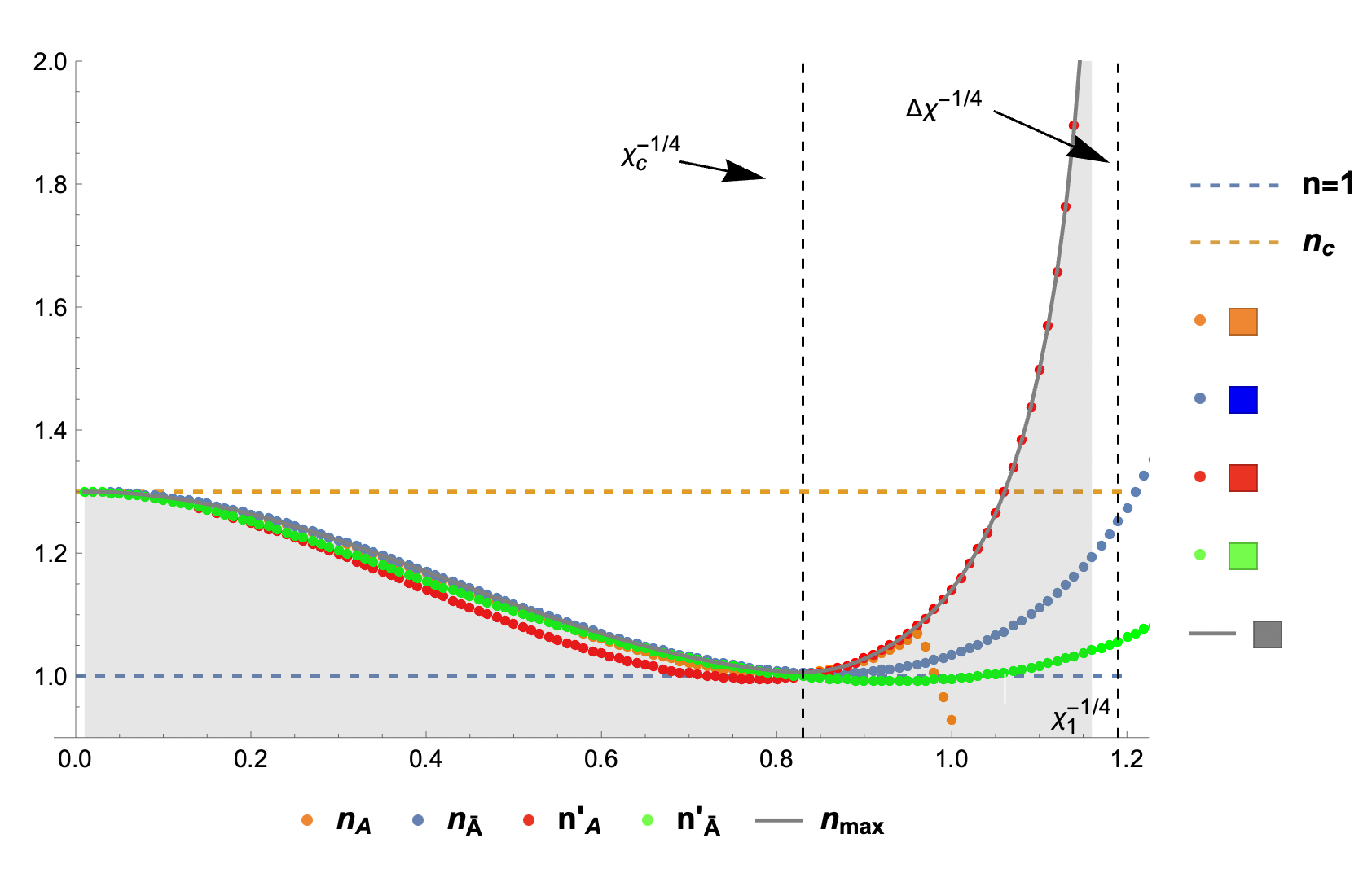}
	\caption{The phase of $n_{cut}(\chi_1)$ with $n_c=1.3$.The individual curves $n_{A,\bar{A}}$ represent the second order instabilities along $A$ and $\bar{A}$ respectively; and $n'_{A,\bar{A}}$ represent the first order instabilities along $A$ and $\bar{A}$ respectively. The lower cut-off $n_{cut}$ is the maximum among these curves.}
	\label{fig:renyi_phases}
\end{figure}
 
In terms of the small parameter $\epsilon$, we conclude that away from the low temperature gap $\chi_1 \gg \Delta \chi_1 \sim \epsilon$, $n_{cut}$ remains close to 1, i.e. $n_{cut}-1 \lesssim n_c -1\sim \epsilon$; it approaches the closest to 1 with $n_{cut}-1 \sim \epsilon^3$ at $\chi_1 =\chi_c \approx 4$. We illustrate these in figure (\ref{fig:renyi_phases}), which shows the phases of $n_{cut}(\chi_1)$ according to the numerically computed values of $n_{A,\bar{A}}$ and $n'_{A,\bar{A}}$ for an explicit choice of $n_c =\sqrt{q_{\bar{A}}/q_A}$.  

\subsection{Other glued solutions}
Having understood the range of validity for the glued BTZ solution (\ref{eq:glued_BTZ_charges},\ref{eq:glued_BTZ_GGE_q5}), we address the remaining issue concerning its status as the dominant saddle-point for $\text{Tr} \rho_A^n$. In practice, this amounts to asking whether there exist other glued solutions to the gluing/matching conditions yielding lower Renyi entropies. An obvious alternative glued solution is the other branch of (\ref{eq:2_branches}): 
\be
q_A= q_1 + \left(1/f-1\right)^{1/2}\sqrt{\Delta q_3},\;\; q_{\bar{A}} = q_1 - \left(1/f-1\right)^{-1/2}\sqrt{\Delta q_3}
\ee
For positive temperature $\beta>0$, the remaining GGE parameters are still given by (\ref{eq:glued_BTZ_GGE_q5}), but for this branch we have $q_A>q_{\bar{A}}$. In the near-primary regime, after a careful analysis it is revealed that this branch can never be a valid saddle-point of $\text{Tr}\rho_A^n$, despite having positive temperatures. More precisely, it can be checked that for $n\geq 1$ and $q_A>q_{\bar{A}}$, the only possibility requires that $h=q_A$ be a $p=1$ limit BTZ in the GGE $(n\beta,\mu^3_n,\mu^1_n)$; $\bar{h}=q_{\bar{A}}$ be a $p=0$ limit BTZ  in the GGE $(\beta,\mu^3_n,\mu^1_n)$; and both GGEs admit three BTZ solutions. Using the results of the BTZ phase diagram in section (\ref{sec:thermo_onezone}), one can then deduce that this is impossible.   

We now discuss the possibility of glued solutions consisting of more general black hole segments, e.g. one-zone black holes. This requires that we solve the gluing/matching condition by assuming more general KdV charge relations representing one-zone black holes. It is a difficult but in principle doable computation. We will come back to this in the discussion section. For the moment let us observe that as $n$ decreases, for $\chi_1>4$ the second-order instabilities are triggered as soon as the BTZ segments become deformable to nearby one-zone black holes; for $\chi_1<4$ the first-order instabilities are triggered before the BTZ segments become deformable to nearby one-zone black holes. We therefore conclude that for $n>n_{cut}$, the glued BTZ solution does not admit deformations to other glued solutions consisting of nearby one-zone black holes.   

We conclude therefore that for $n>n_{cut}$, the glued BTZ solution (\ref{eq:glued_BTZ_charges},\ref{eq:glued_BTZ_GGE_q5}) is the only glued solution that is valid. We can therefore trust the Renyi entropies (\ref{eq:mixed_renyi}) for $n>n_{cut}$.  Admittedly, the perturbative analysis considers only nearby configurations when arguing for the thermodynamic stability of the BTZ segments and the absence of more general glued solutions. Intuitively, in the near-primary regime $\epsilon = \sqrt{q_3 -q_1^2}/q_1\ll 1$ such considerations are likely to capture the full picture. We leave the task of non-perturbative analysis to the future. This is important for studying more general cases, e.g. ensembles with generic fixed KdV charges that are away from the near-primary regime.  

\subsection{Implications for the entanglement spectrum}
Let us summarize the results in terms of the Renyi entropy. For $\epsilon = \sqrt{q_3-q_1^2}/q_1 \ll 1$ and consider the micro-canonical and mixed ensemble of KdV charges: 
\be 
\rho^{micro}_{q_1,q_3} = \hat{P}_{q_1,q_3},\;\;\;\rho^{\beta}_{q_1,q_3} =\mathcal{N}^{-1} e^{-\beta \hat{Q}_5}\; \hat{P}_{q_1,q_3}
\ee
The Renyi entropy $S_n(A)$ is simply given by:
\be
S_n(A) =\frac{ \frac{\pi}{2G_N}f\sqrt{q_A}+ \frac{\pi}{2G_N}n(1-f)\sqrt{q_{\bar{A}}}-n S_{th}}{1-n},\;\;\;q_A = q_1\left(1-\epsilon\sqrt{f^{-1}-1} \right) 
\ee
This result is valid for Renyi indices $n> n_{cut}$, where the lower cut-off $n_{cut}=\sqrt{q_{\bar{A}}/q_A}$ in $\rho^{micro}_{q_1,q_3}$, and depends on the rescaled temperature $\chi_1 = \pi T/q_{\bar{A}}^{5/2}$ according to the phases illustrated in figure (\ref{fig:renyi_phases}) for $\rho^{\beta}_{q_1,q_3}$.  For $n<n_{cut}$, the only knowledge is its value at $n=1$, given by the von Neumann entropy:  
\be 
S_1(A) = S_{vN}(A) =  f S_{th}
\ee
The interpolation from $n=1$ to $n\geq n_{cut}$ depends on the resolutions of the instabilities discussed in section (\ref{sec:instability}). They are beyond the scope of this work, and we leave its discussions to section (\ref{sec:discussion}).  

\begin{figure}[ht]
	\centering
	\includegraphics[width=0.49\linewidth]{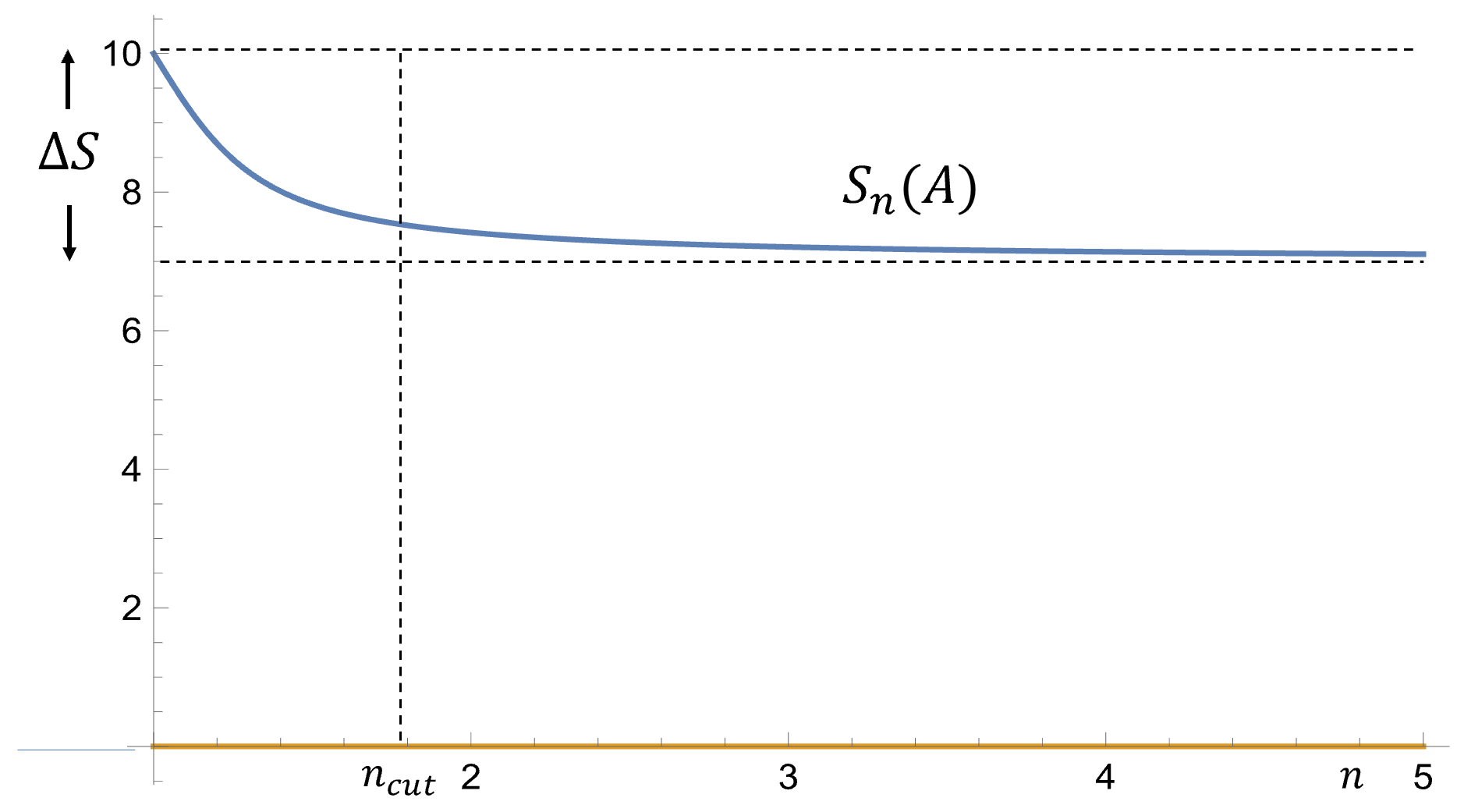}
     \includegraphics[width=0.49\linewidth]{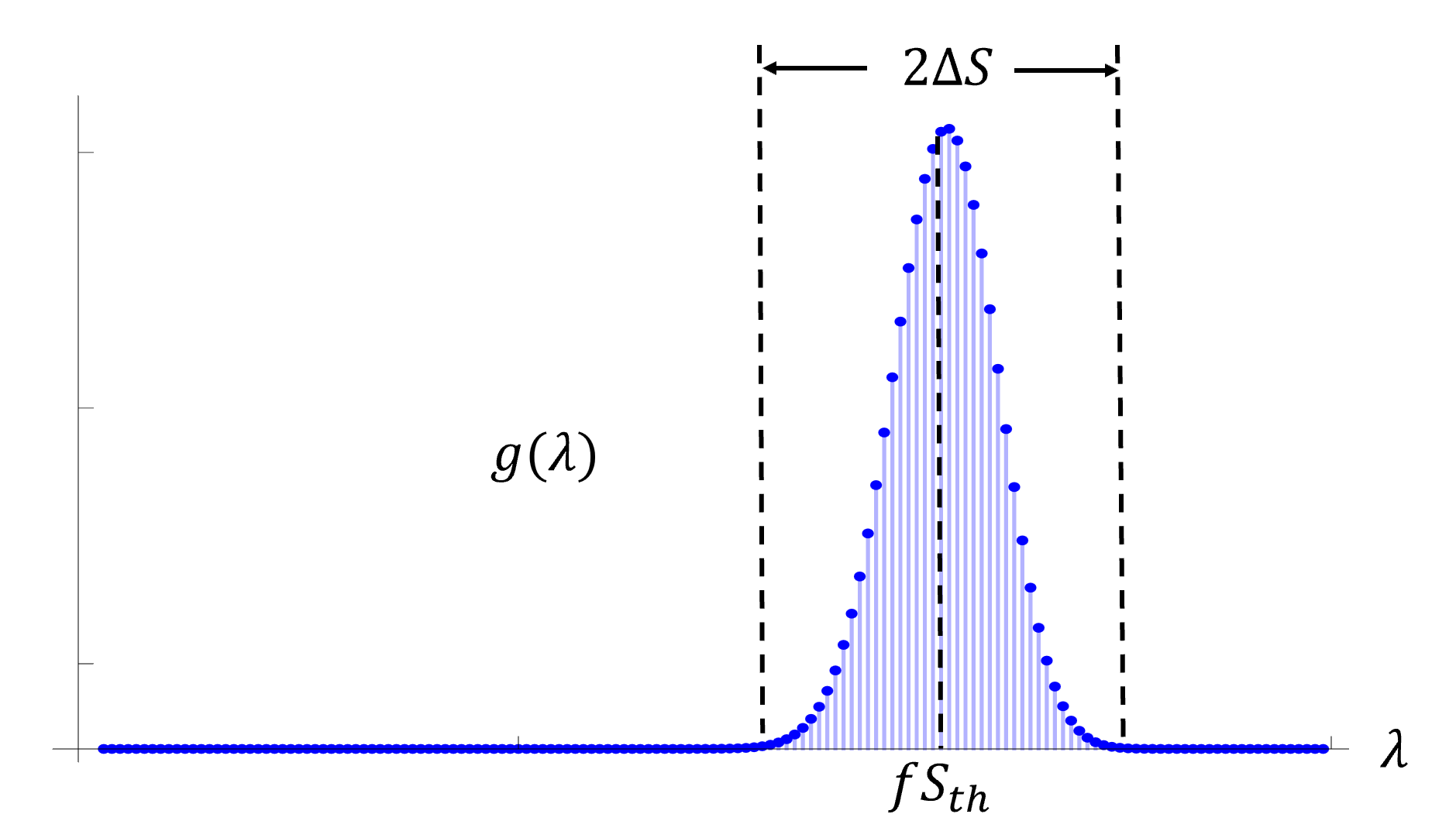}
	\caption{Left: general feature of the Renyi entropy $S_n(A)$; right: general features of the entanglement spectral density $g(\lambda)$ implied by the left.}
	\label{fig:renyi_to_spectrum}
\end{figure}

Now we explore some implications. An important aspect of the entanglement properties regarding the reduced density matrix $\rho_A = \text{Tr}_{\bar{A}} \rho^{\beta}_{q_1,q_3}$ is its entanglement spectral density $g(\lambda)$. It is related to the Renyi entropies via Laplace and inverse Laplace transformations: 
\bea\label{eq:inverse_laplace}
\text{Tr} \rho^n_A &=& \int d\lambda\; g(\lambda) e^{-n\lambda} =e^{-(n-1)S_n(A)},\;\;\rho_A = \int d\lambda\; e^{-\lambda}|\lambda \rangle \langle \lambda|\nonumber\\
g(\lambda)&=& \frac{1}{2\pi i}\int^{\Gamma+i\infty}_{\Gamma-i\infty} dn' \;e^{n'\lambda} \;e^{(1-n')S_{n'}(A)}
\eea
With only the partial knowledge of $S_n(A)$ for $n>n_{cut}$, it is difficult to perform the inverse Laplace transform explicitly. We instead aim at deducing features of the entanglement spectral density $g(\lambda)$ that could consistently reproduce the qualitative behaviors of the Renyi entropies. We focus on those that are relevant at $\epsilon \ll 1$ for $\rho^{micro}_{q_1,q_3}$ and the high temperature phase of $\rho^{\beta}_{q_1,q_3}$: 
\begin{itemize}
\item  As the Renyi-index $n$ varies between $[1,\infty]$, the value of the Renyi entropy is bounded by the asymptotic values in a window of width: 
\be
\Delta S =  S_1(A) - S_{\infty}(A) \propto  \epsilon f S_{th} 
\ee
The most prominent feature is that $\Delta S$ shrinks with vanishing $\epsilon$. 
\item The Renyi entropy approaches a constant, i.e. becomes independent of $n$, for sufficiently large $n-1 \gg \delta n_{cut}$. The most prominent feature is that $\delta n_{cut}$ also shrinks with vanishing $\epsilon$, see Figure (\ref{fig:renyi_to_spectrum}).  
\end{itemize}  

In order to be consistent with these features, we propose that the entanglement spectral density are characterized by a bounded support:
\be
g(\lambda) = 
\begin{cases} 
g(\lambda), \;\;f S_{th}-\Delta S \leq \lambda  \leq f S_{th}+\Delta S\\
0,\;\;\;\;\text{otherwise}\\
\end{cases}
\ee
with vanishing densities towards the edges of the window, see Figure (\ref{fig:renyi_to_spectrum}). Correspondingly, the most prominent feature is that the width of the spectral support coincides with $\Delta S$ and thus also shrinks with vanishing $\epsilon$. In the appendix (\ref{app:ent_spectrum}), we demonstrate this in a toy model expression of $S_n(A)$ that exhibits similar features. 

We end this section with the following comment. It is tempting to extrapolate this observation to the actual primary states with $\epsilon \to 0$, towards which the entanglement spectral density collapse to a single delta functional peak: 
\be
g(\lambda) = e^{S_0}\delta(\lambda-S_0) 
\ee
and resulting in an $n$-independent Renyi entropy for all $n\geq 1$: 
\be
S_n(A) = S_{vn}(A),\;\;\;n\geq 1 
\ee
In other words, the extrapolation to the primary states at $\epsilon \to 0$ yields a flat entanglement spectrum. States whose entanglement properties exhibit this feature are important in understanding the backbones of AdS/CFT. For example, they characterize some tensor network models of holography \cite{TN1,TN2}; they are also related to the so-called fixed area states that underly the effective configuration space of quantum gravity \cite{fixedarea1,fixedarea2,fixedarea3}. 

\section{Discussions}\label{sec:discussion}
In the final section, we first summarize the main points of the paper. After that we discuss some remaining issues, along which potential outlooks for future investigations will be suggested. 

\subsection{Summary}
In this paper, we discussed the computation of holographic Renyi entropies for ensembles with fixed KdV charges, and used the results to explore the underlying entanglement properties. This is relevant to the question of subsystem ETH in holographic 2d CFTs. The computation utilizes two ingredients: the cosmic-brane prescription, which is applicable to generic holographic states; and the gluing construction, which is an approximation scheme to solve the cosmic-brane back-reaction. The gluing construction was proposed in \cite{Dong:2018}, and we extended it to cases with fixed KdV charges. As an approximation scheme, it is effective when computing the Renyi entropies at the leading order in the high KdV charge density limit. This is the limit our computation focused on in this work. 

To be explicit, we performed the computation on cases with the first two KdV charges fixed to $\langle \mathcal{Q}_1\rangle = q_1$ and $\langle \mathcal{Q}_3\rangle = q_3>q_1^2$. We first considered the micro-canonical ensemble: 
\be
\rho^{micro}_{q_1,q_3} = \mathcal{N}^{-1} \hat{P}_{q_1,q_3}
\ee 
and subsequently extended to mixed ensembles decorated with a temperature $\beta$ for the next KdV charge $\mathcal{Q}_5$:
\be\label{eq:mmGGE_2}
\rho^{\beta}_{q_1,q_3} = \mathcal{N}^{-1} \hat{P}_{q_1,q_3}\;e^{-\beta \hat{Q}_5} 
\ee
The gluing construction involves finding segments of black hole geometries that solve a set of gluing/matching conditions. For our cases these black holes carry KdV charges in $\text{AdS}_3/\text{CFT}_2$. They are described by the so-called finite-zone solutions, of which the BTZ black holes is a special class with zero-zone. We systematically surveyed the thermodynamic properties of the BTZ and one-zone black holes in the relevant ensembles. Based on the results, we focused on the glued-solutions in the near-primary regime between the fixed KdV charges: 
\be
\epsilon = \sqrt{q_3/q_1^2-1} \ll 1 
\ee
We found that for sufficiently large Renyi index $n>n_{cut}$, the dominant glued solution takes the form of two segments of BTZ black holes of $\mathcal{Q}_1$ charge densities:
\be 
q_A = q_1 \left(1-\epsilon\sqrt{f^{-1}-1}\right),\;\;q_{\bar{A}}=q_1 \left(1+\frac{\epsilon}{\sqrt{f^{-1}-1}}\right)
\ee
For $n>n_{cut}$, the Renyi entropy $S_n(A)$ is equal to: 
\be 
S_n(A) = \frac{ \frac{\pi}{2G_N}f\sqrt{q_A}+ \frac{\pi}{2G_N}n(1-f)\sqrt{q_{\bar{A}}}-n S_{th}}{1-n}
\ee
This features an $n$-independent refined Renyi entropy: 
\be
\tilde{S}_n(A) = \frac{f\pi\sqrt{q_A}}{2G_N}
\ee
The lower cut-off $n_{cut}$ for the Renyi index corresponds to when the glued solution becomes unphysical. For the micro-canonical ensemble it is given by: 
\be 
n_{cut}=n_c = \sqrt{\frac{q_{\bar{A}}}{q_A}} = 1 + \frac{\epsilon}{2\sqrt{f(1-f)}} +...
\ee
below which the BTZ segments have negative temperatures. For the mixed ensemble $n_{cut}$ depends on both $n_c$ and the temperature through the combination $\chi_1=\pi T/q_{\bar{A}}^{5/2}$. It corresponds to when the BTZ segments become unstable in the corresponding GGE. We found that $n_{cut}$ is also close to 1, i.e. $n_{cut}-1\lesssim \epsilon$, for sufficiently high temperatures satisfying $\chi_1 \gtrsim \chi_c \approx 4$. The general features of the Renyi entropy imply that the underlying entanglement spectral density $g(\lambda)$ is characterized by a bounded support: 
\be 
f S_{th}-\Delta S \leq \lambda \leq f S_{th} + \Delta S
\ee
The extrapolation to $\epsilon \to 0$ then reveals a flat entanglement spectrum, which is reminiscent of the fixed-area states in AdS/CFT. 

\subsection{Primary states v.s. fixed-area states}
The most prominent feature of the holographic Renyi entropy at fixed KdV charges $(q_1,q_3)$ is the $n$-independence of the refined Renyi entropy $\tilde{S}_n(A)$ for $n>n_{cut}$, where $n_{cut}\to 1$ in the $\epsilon \to 0$ limit towards primary states. We can interpret this behavior as describing the restricted nature of the gravitational back-reaction in the bulk dual of the mixed ensemble (\ref{eq:mmGGE_2}), upon the insertion of cosmic-branes. It can be contrasted with that of ordinary BTZ black holes representing the micro-canonical ensembles.  Intuitively the restriction is a result of the additional conservation law imposed on the KdV charges, which then affects the gravitational dynamics in $\text{AdS}_3/\text{CFT}_2$.  

Taking $\epsilon \to 0$  then leads to a flat entanglement spectrum across any finite interval $A$. The Renyi entropy $S_n(A)$ is equal to the von-Neumann entropy for all $n\geq 1$. In these states, the insertion of cosmic-branes produces no effect on the minimal surface area. This is the defining character of the fixed-area states that encode super-selection sectors of the bulk configuration space \cite{fixedarea1}. In other words, the gravitational back-reaction is restricted to the maximal extent -- it appears to be ``frozen" in the $\epsilon\to 0$ limit. One can understand this as follows. The saturation of $q_3 \geq q_1^2$ automatically implies the saturation of infinitely many relations $q_{2k-1}\geq q_1^k$ among the KdV charges. The states in the $\epsilon \to 0$ limit is therefore implicitly defined by infinitely many conservation laws restricting its gravitational interaction with cosmic-branes. The flat entanglement spectrum may be a consequence of this. We remark that in the $\epsilon \to 0$ limit, the metric of the original bulk dual is the same as an ordinary BTZ black hole. Its characterizations from $\epsilon \to 0$ are encoded in the response to cosmic-brane insertions. 

On the other hand, from the CFT side the computation for the Renyi entropy in primary states $|h\rangle = \mathcal{O}_h |\Omega\rangle$ has been performed in \cite{Wang:2018}. It focused on the same limit of our interest, i.e. the leading order results for a finite interval in the high energy density limit $h/c \propto q_1\gg 1$ while sending $c\to \infty$.  For pure states, we have to restrict to subsystems smaller than half of the total size, i.e. $f<1/2$. In 2d CFTs, the Renyi entropy $S_n(A)$ is related to the correlation function $\langle \mathcal{O}_h\sigma_n \sigma_n \mathcal{O}_h\rangle$ in the orbifold CFT, where $\sigma_n$ is the twist operator. The computation was done via the method of monodromy, which computes the Virasoro vacuum block contribution to the correlation function in the $c\to \infty$ limit. This is essentially computing the gravitation back-reaction. The monodromy problem was solved in the high energy limit using the WKB approximation. The leading order result for $S_n(A)$ is $n$-independent and hence implies a flat entanglement spectrum. We therefore had an independent computation that verify the extrapolation directly for the primary states in 2d CFTs. We perceive this as in support of the subsystem ETH for the primary states according to their higher KdV charges. It is reasonable to expect that our results can be extended to all near-primary states in the high density limit, even for $\epsilon \lesssim c^{-1}$. This is indeed the case based on the results in \cite{KdVETHgeneral}.  

We clarify by emphasizing that in the $\epsilon \to 0$ limit, the fixed-area property only describes the leading order behavior of the Renyi entropy, in particular the part scaling with the total volume, or equivalently the total charge of the state. It is not clear whether the sub-leading contributions exhibit such properties. In the future, it is interesting to extend the analysis to subleading orders. Besides, recall that at finite $\epsilon$ there exists a critical temperature $\chi_1 = \chi_c$ at which $n_{cut}-1$ is further suppressed to order $\mathcal{O}(\epsilon^3)$, extending the range of $n$-independence for $\tilde{S}_n(A)$ to the maximum. In the future, it is interesting to understand what underlies this.  

\subsection{Beyond instabilities}
For the mixed ensembles $\rho^{\beta}_{q_1,q_3}$ we had based our analysis on identifying the instabilities of the glued BTZ solutions. We now discuss the nature of these instabilities, which may shed light on what the back-reacted solution becomes for $n<n_{cut}$. We have classified the relevant instabilities into the first order and second order types. We first discuss the second order instabilities. They are identified by a BTZ black segment, say along $A$, becomes unstable against nearby one-zone black holes in the corresponding GGE. It is reasonable to speculate that by crossing $n_{cut}$ of this nature, the back-reacted geometry takes the form of glued finite-zone black holes, at least along segment $A$. For the sake of discussion let us assume it is a one-zone black holes characterized by $p>0$, where we recall: 
\be
p= \frac{\lambda_3-\lambda_2}{\lambda_3-\lambda_1}
\ee 
The free energy functional is expected to vary continuously, thus the lowest energy configuration also changes continuously from $p=0$ to $p>0$ across $n=n_{cut}$. As a result, we expect the properties of the glued solution to change continuously across the transition, with $p=0$ on one side and $p>0$ on the other. In particular, the Renyi entropy $S_n(A)$, which represents the free energy of the partition function $Z(\Sigma^n_A)$, and the refined Renyi entropy $\tilde{S}_n(A)$, which represents the derivative of the free energy, are both continuous across the transition. This is reminiscent of a second order phase transition. The one-zone parameter $p>0$ can then serve as an order parameter. Computing glued solutions of this nature is in principle tractable, we leave it for future investigations. 

Next we discuss the first order instabilities. They are characterized by one of the BTZ black segments, say with $\mathcal{Q}_1$ charge density $h$ along $A$, switching dominance with another BTZ saddle of $\mathcal{Q}_1$ charge density $h'$ in the corresponding GGE. We emphasize that it does not lead to a first-order phase transition between the two BTZ segments of charge density $h$ and $h'$ -- this violates the matching condition on the total KdV charges. As a result, it is unclear what the bulk saddle of $Z(\Sigma^n_A)$ becomes for $n<n_{cut}$. Recall that $n_{cut}$ is dictated by the first order instability for lower temperatures $\chi_1\lesssim \mathcal{O}(1)$. This is roughly the temperature regime that $\rho^\beta_{q_1,q_3}$ ceases to have a well-defined black hole dual, see (\ref{eq:T_bound}).  The difficulty for finding the glued solution below $n<n_{cut}$ may be a revelation of this fact, analogous to the discussion in section (\ref{sec:glued_BTZ}) regarding $\rho^{micro}_{q_1,q_3}$ for $n<n_c$. It is found in the companion paper \cite{KdVETHgeneral} that at large $c$ and for $n<n_{cut}$, the Renyi entropy in $\rho^{micro}_{q_1,q_3}$ is simply given by that of the ordinary micro-canonical ensemble at the leading order in $1/c$. The bulk implication of this remains unclear. We conjecture that below $n<n_{cut}$, the Renyi entropy can still be computed by the gluing construction, whose validity extends beyond holography, see appendix (\ref{sec:gluing_derivation}). However, the segment of the glued solution can no longer be described by well-defined gravitational saddle-points. The original ensemble $\rho^{\beta}_{q_1,q_3}$ may be described similarly. We leave exploring these possibilities for the future. 

\subsection{Fixing more KdV charges}\label{sec:more}
In this paper we have studied explicitly the holographic Renyi entropies for ensembles with fixed $\langle \mathcal{Q}_1\rangle$ and $\langle \mathcal{Q}_3\rangle$. A natural follow up question is what happens for ensembles with more KdV charges fixed? In particular, how much of the qualitative features may be preserved as we fix more and more KdV charges? Without explicitly performing these computations it is difficult to give concrete answers; we instead discuss some plausible features of the computations based on their general structures. 

The most important question concerns whether the refined holographic Renyi entropy remain $n$-independent, at least in some interval, in ensembles with more KdV charges fixed. To this end, let us first extract the main reason driving behind this. In the case of the glued BTZ solutions, it comes from the fact that the matching and gluing conditions are solved separately. The charge densities of the BTZ segments are fixed from the matching conditions alone, which is independent of $n$, and this determines the refined Renyi entropy; the $n$-dependence is encoded in the gluing conditions, which determine the GGE parameters, but do not affect the refined Renyi entropy. This is to be contrasted with the computation of the ordinary micro-canonical ensembles in \cite{Dong:2018}, in which the gluing/matching condition can only be solved simultaneously, resulting in an $n$-dependent refined Renyi entropy.  

Let us imagine going one step further and solving the gluing/matching condition for fixing the first three KdV charges: 
\be
f \langle \mathcal{Q}_{2k-1}\rangle_{n\beta,\mu^3,\mu^1} + f\langle \mathcal{Q}_{2k-1}\rangle_{\beta,\mu^3,\mu^1} = q_{2k-1},\;\;\;\;k=1,2,3  
\ee
By choosing to work with GGEs of three chemical potentials, we are computing the Renyi entropy in the micro-canonical ensemble $\rho^{micro}_{q_1,q_3,q_5}$.  In this case, the matching condition can no longer be satisfied by gluing two BTZ black hole segments. The reason is that for each BTZ black hole segment, the KdV charges depend only on one parameter. A glued BTZ solution therefore has two independent parameters, which is over-determined to satisfy three matching conditions. We can naturally relax one of the black hole segments to be a one-zone black hole, whose KdV charges depend on three parameters, namely the zone-parameters $(\lambda_1,\lambda_2,\lambda_3)$. In this case the glued solution consists of a BTZ segment and a one-zone segment. It has four independent parameters which is sufficient for satisfying three matching conditions. It is in fact under-determined. However, what matters is that for any choice of such glued solutions, it is always possible to find a set of GGEs that satisfy the gluing conditions. There are in total three equations -- one from the BTZ segment and two from the one-zone segment, for the three independent parameters $\lbrace\beta,\mu^3,\mu^1\rbrace$. As a result, similar to the glued BTZ solution, the matching and gluing conditions are solved separately. Due to the under-determinacy of the procedure, we would obtain a class of glued solutions, each giving a refined Renyi entropy $\tilde{S}_n(A)$ that is $n$-independent. 

From this example, we conjecture that the separation between solving the matching and gluing conditions is likely to remain when more KdV charges are fixed. Whether this eventually leads to refined Renyi entropies that are $n$-independent (at least piece-wise) would require further studies. For example, with a class of glued solutions just described, there could be a few possibilities regarding the optimal one as $n$ is varied. It may undergo a series of phase transitions; or it may change continuously with $n$. We should point out that if we fix four KdV charge $\langle \mathcal{Q}_{2k-1}\rangle=q_{2k-1},k=1,...,4$, there is a unique glued solution consisting of a one-zone and a BTZ segment satisfying the matching conditions, which is more analogous to the glued BTZ solution. So the answer may also depend on whether an even/odd number of KdV charges are fixed.  We leave these for future investigations.   

\section*{Acknowledgments}
We thank Xi Dong, Sotaro Sugishita, Jieqiang Wu, and Long Zhang for useful discussions. L.C and H.W are supported by National Science Foundation of China (NSFC) grant no. 12175238.  JT is supported by  the National Youth Fund No.12105289.
AD is supported by the NSF under  grant PHY-2310426. This work was performed in part at Aspen Center for Physics, which is supported by National Science Foundation grant PHY-2210452.

\appendix

\section{Conditions (\ref{eq:smoothness}) v.s. isolated BTZ black hole }\label{sec:smooth_BTZ}
We address a potential paradox regarding the isolated BTZs with real zone parameters $(h,w)$. As pointed out, they emerge as limits of one-zone black holes that fail to satisfy the smoothness and physical conditions, yet are themselves perfectly smooth and physical. It appears in these cases that the conditions (\ref{eq:smoothness}) cannot be extrapolated to at least one of the BTZ limits, i.e. $p\to 0$ or $p\to 1$. To understand this let us quote some of the details in \cite{Dymarsky:2020} when deriving the first condition in (\ref{eq:smoothness}), which comes from requiring that $f$ be positive definite. For one-zone solutions in (\ref{GGE3}), recall that we have: 
\bea\label{eq:smoothness_onezone}
&&f(\varphi) = 2(\mu_3+s_1) \left(u(\varphi)-s_1\right)>0,\;\;s_1 = u_1+u_2+u_3\nonumber\\
&&u_1 = 4(\lambda_1+\lambda_2-\lambda_3),\;u_2 = 4(\lambda_1-\lambda_2+\lambda_3),\;\;u_3 = 4(-\lambda_1+\lambda_2+\lambda_3)
\eea
In addition, it can be derived that requiring $u_0>0$ forces the zone parameters to be one of the following arrangements: 
\be\label{eq:onezone_constraint_1}
\lbrace\lambda_3\geq \lambda_2\geq \lambda_1>0\rbrace\;\;\;\;   \text{or}\;\;\;\; \lbrace\lambda_3>0>\lambda_2\geq \lambda_1\rbrace
\ee
For any one-zone solution away from the $p=1$ limit, i.e. $\lambda_1=\lambda_2$, the profile $u(\varphi)$ of the solution oscillates between the interval $[u_1,u_2]$. This also includes the $p\to 0$ limit, i.e. $\lambda_2 = \lambda_3$, for which $u_1 = u_2$ and $u(\varphi)$ is correctly constrained to be constant. For these solution, one can obtain the maximum of $f(\varphi)$ by plugging in $u(\varphi)=u_{1,2}$ depending on the sign of $(\mu_3+s_1)$, and positive definiteness of $f(\varphi)$ imposes one of the following constraints: 
\bea\label{eq:onezone_constraint_2}
\lbrace \mu_3<-s_1,\;\lambda_2>0 \rbrace \;\;\;\;\text{or}\;\;\;\;\lbrace \mu_3>-s_1,\;\lambda_3<0\rbrace
\eea
Combining both sets of conditions (\ref{eq:onezone_constraint_1}) and (\ref{eq:onezone_constraint_2}) then gives part of the smoothness and physical conditions (\ref{eq:smoothness}). 

On the other hand, the BTZ black holes from the $p\to 1$ limit, i.e. $\lambda_1=\lambda_2$ poses an exception to this argument. The constant profile $u(\varphi)=u_2$ does not oscillate between $[u_1,u_2]$ despite $u_1<u_2$. This then alters the analysis of (\ref{eq:onezone_constraint_2}), and yields instead the condition: 
\be\label{eq:onezone_constraint_3}
\lbrace \mu_3<-s_1,\;\lambda_2>0 \rbrace \;\;\;\;\text{or}\;\;\;\;\lbrace \mu_3>-s_1,\;\lambda_2<0\rbrace 
\ee
which brings the following new possibility to satisfy the smoothness and physical condition: 
\be
\mu_3>-s_1,\;\;\lambda_3>0>\lambda_2=\lambda_1 
\ee
This corresponds to and thus characterizes the isolated BTZ black hole with real-valued zone-parameters. As discussed before, physically the $p\to 0$ limit is approached by profiles $u(\varphi)$ with diminishing oscillating amplitudes; while the $p\to 1$ limit is approached by diminishing frequency $k \to 0$ with a potentially large amplitude. This explains the ``jump" in the smoothness and physical condition away from isolated BTZ black holes of the $p\to 1$ limit.

\section{Phases of BTZ black holes}\label{app:ReBTZ}
\renewcommand{\theequation}{A.\arabic{equation}}
\setcounter{equation}{0}
In this appendix we supplement some details in deriving the phase diagram summarized in the table (\ref{tb:phases}). We have written down in section (\ref{sec:thermo_onezone}) the inequalities among $(h,T,\mu_3)$ characterizing the properties of the underlying BTZ black hole. Next we organize these inequalities into phases for $\mu_3$ at fixed $(h,T)$. To this end, let us first define the following quantities: 
\bea
\zeta_1 &=& - \left(\frac{\pi T}{2 h^{3/2}}\right)-3h,\;\; \zeta_2=-5h\nonumber\\
\zeta_3 &=& -\left(2\sqrt{2}-1\right)h - \frac{\pi T}{2(\sqrt{2}-1) h^{3/2}},\;\;\zeta_4 = -\left(\frac{4\pi T}{\sqrt{h}}\right)^{1/2}-h
\eea 
Among them, one can check that we always have $\zeta_1\leq \zeta_4,\;\;\zeta_3\leq \zeta_4$. Then we can assemble and arrange the inequalities into the following phases as $\mu_3$ is varied:
\begin{itemize}
\item For $\mu_3<\zeta_1$, the BTZ corresponds to a local maximum of $\mathcal{F}_{BTZ}$ and thus cannot be considered as the thermodynamically dominant saddle-point of the GGE $(\beta,\mu_1,\mu_3)$. 
\item For $\zeta_1 <\mu_3<\text{min}\left\lbrace\zeta_2, \zeta_4\right\rbrace$, both branches $(h,w^{\pm})$ are deformable as the $p\to 0$ limit of one-zone black holes.
\item For $\text{max}\left\lbrace\zeta_1,\;\zeta_2\right\rbrace<\mu_3<\zeta_4$, both branches $(h,w^{\pm})$ could be deformable as the $p\to 1$ limit of one-zone black holes if they further satisfy: 
\be
w^{\pm}>\left(\sqrt{2}-1\right)h 
\ee
It turns out that this depends on the sign of: 
\be 
\Delta_p = \left(\sqrt{2}-1\right)h-\left(\frac{\pi T}{4\sqrt{h}}\right)^{1/2}
\ee
If $\Delta_p>0$, only the $w^+$ branches is likely to be deformable, and it is so for:
\be
\mu_3 < \zeta_3 
\ee
If $\Delta_p<0$, the $w^{+}$ branch is automatically deformable, and the other branch $w^{-}$ is also deformable for:
\be
\zeta_3 <\mu_3 <\zeta_4
\ee
\item For $\mu_3>\zeta_3$ in the case of $\Delta_p>0$ and $\mu_3>\zeta_4$ in the case of $\Delta_p<0$, the BTZ is isolated.  
\end{itemize}
Some of the phases could be absent if the corresponding window closes. This depends on $(h,T)$. To facilitate further analysis, we define the re-scaled parameters:
\bea
\chi_1 &=& \left(\frac{\pi T}{h^{5/2}}\right)>0,\;\;\chi_2 = \left(\frac{\mu_3}{h}\right),\;\;\tilde{\zeta}_1 = -\frac{\chi_1}{2}-3\nonumber\\
\tilde{\zeta}_2&=&-5,\;\;\tilde{\zeta}_3 = -\left(2\sqrt{2}-1\right)-\frac{\chi_1}{2\left(\sqrt{2}-1\right)},\;\;\tilde{\zeta}_4=-2\sqrt{\chi_1}-1
\eea
We then find the following intervals for $\chi_1$ defined by $(\alpha_1,\alpha_2,\alpha_3)$ of values: 
\be 
 \alpha_1 = 4\left(\sqrt{2}-1\right)^2,\;\;\alpha_2=4\left(\sqrt{2}-1\right),\;\;\alpha_3 =4
\ee 
that characterize distinct phase structures as the $\chi_2$ is varied:
\begin{itemize}
\item $\chi_1\in [\alpha_3,\;\infty]$: in this range we have that:
\be
\zeta_3<\tilde{\zeta}_1<\tilde{\zeta}_4<\tilde{\zeta}_2,\;\;\Delta_p<0
\ee 
with the following phases:
\be
\chi_2\in 
\begin{cases}
[\tilde{\zeta}_4,\infty]:\;\;\;\text{both $(h,w^{\pm})$ are isolated}\\
[\tilde{\zeta}_1,\tilde{\zeta}_4]:\;\;\; \text{both $(h,w^{\pm})$ are deformable as the $p\to 0$ limit}\\
[-\infty,\tilde{\zeta}_1]:\;\;\;\text{$h$ is a local maximum of $\mathcal{F}_{BTZ}$}
\end{cases}
\ee

\item $\chi_1\in [\alpha_2,\;\alpha_3]$: in this range we have that: 
\be
\left(\tilde{\zeta}_2,\;\tilde{\zeta}_3\right)<\tilde{\zeta}_1<\tilde{\zeta}_4,\;\;\Delta_p<0,
\ee 
with the following phases:
\be
\chi_2\in 
\begin{cases}
[\tilde{\zeta}_4,\infty]:\;\;\;\text{both $(h,w^{\pm})$ are isolated}\\
[\tilde{\zeta}_1,\tilde{\zeta}_4]:\;\;\; \text{both $(h,w^{\pm})$ are deformable as the $p\to 1$ limit}\\
[-\infty,\tilde{\zeta}_1]:\;\;\;\text{$h$ is a local maximum of $\mathcal{F}_{BTZ}$}
\end{cases}
\ee

\item $\chi_1\in [\alpha_1,\;\alpha_2]$: in this range we have that: 
\be
\tilde{\zeta}_2<\tilde{\zeta}_1<\zeta_3<\tilde{\zeta}_4,\;\;\Delta_p<0
\ee 
with the following phases:
\be
\chi_2\in 
\begin{cases}
[\tilde{\zeta}_4,\infty]:\;\;\;\text{both $(h,w^{\pm})$ are isolated}\\
[\tilde{\zeta}_3,\tilde{\zeta}_4]:\;\;\; \text{both $(h,w^{\pm})$ are deformable as the $p\to 1$ limit}\\
[\tilde{\zeta}_1,\tilde{\zeta}_3]:\;\;\; \text{only $(h,w^+)$ is deformable as the $p\to 1$ limit}\\
[-\infty,\tilde{\zeta}_1]:\;\;\;\text{$h$ is a local maximum of $\mathcal{F}_{BTZ}$}
\end{cases}
\ee

\item $\chi_1\in [0,\; \alpha_1]$: in this range we have that: 
\be
\tilde{\zeta}_2<\tilde{\zeta}_1<\tilde{\zeta}_3<\tilde{\zeta}_4,\;\;\Delta_p>0
\ee 
with the following phases:
\be
\chi_2\in 
\begin{cases}
[\tilde{\zeta}_3,\infty]:\;\;\;\text{both $(h,w^{\pm})$ are isolated}\\
[\tilde{\zeta}_1,\tilde{\zeta}_3]:\;\;\; \text{only $(h,w^+)$ is deformable as the $p\to 1$ limit}\\
[-\infty,\tilde{\zeta}_1]:\;\;\;\text{$h$ is a local maximum of $\mathcal{F}_{BTZ}$}
\end{cases}
\ee
\end{itemize}

\section{Additional support for the gluing construction (\ref{eq:canonical_match},\;\;\ref{eq:micro_match})}\label{sec:gluing_derivation}
We supplement additional support for the gluing constructions in section (\ref{sec:renyi}) based on the following ansatz for the density matrix of the canonical ensemble: 
\be \label{eq:CE_ansatz}
\rho^\beta = \mathcal{N}^{-1}\sum_{ij} e^{-\beta (E_i+\bar{E}_j)} |E_i\rangle_A \otimes |\bar{E}_j\rangle_{\bar{A}}\; \langle E_i|_A \otimes \langle \bar{E}_j|_{\bar{A}}
\ee
This can be derived from the following chaotic ansatz for the energy eigenstates: 
\be 
|E\rangle = \sum_{ij} c_{ij} |E_i\rangle_A \otimes |E_j\rangle_{\bar{A}}
\ee
where $c_{ij}$ are random variables satisfy: 
\be
\overline{c_{ij}c_{i'j'}} = \delta_{ii'}\delta_{jj'} 
\ee
from which (\ref{eq:CE_ansatz}) can be obtained as the statistical average. In these ansatz, $|E\rangle_A$ and $|\bar{E}\rangle_{\bar{A}}$ are eigenstates of the subsystem Hamiltonians whose sum is approximately the total Hamiltonian: 
\be 
H_A|E\rangle_A = E|E\rangle_A ,\;\;\;H_{\bar{A}}|\bar{E}\rangle_{\bar{A}} = \bar{E}|\bar{E}\rangle_{\bar{A}},\;\;\;H\approx H_A\otimes \mathds{1}_A + \mathds{1}_A \otimes H_{\bar{A}}
\ee
This property encodes the assumptions that we are considering a subsystem $A$ of finite fraction $f$ in the high density limit. Based on (\ref{eq:CE_ansatz}) it is straight-forward to first write down the reduced density matrix across the subsystem A: 
\be
\rho^\beta_A = \text{Tr}_{\bar{A}}\; \rho^{\beta} \propto Z_{\bar{A}}(\beta) \sum_{i} e^{-\beta E_i} |E_i\rangle\; \langle E_i| 
\ee
where $Z_{\bar{A}}(\beta)$ is the partition function of the subsystem Hamiltonian $H_{\bar{A}}$. The corresponding trace giving the Renyi entropy of the canonical ensemble can then computed by: 
\be 
\text{Tr}_A \left(\rho^\beta_A\right)^n = Z_A(n \beta) Z_{\bar{A}}(\beta)^n
\ee
This is equivalent to the gluing construction for the canonical ensemble Renyi entropy, which only enforces the gluing condition:
\be
\beta_A = n \beta_{\bar{A}} =\beta
\ee
To obtain the micro-canonical counter-part, we first perform an inverse laplace transform of the ansatz for the canonical ensemble (\ref{eq:CE_ansatz}): 
\be
\rho^E = \oint_{\Gamma} d\beta\; e^{\beta E}\; \rho^\beta 
\ee 
where $\Gamma$ is the corresponding Bromwich contour whose form will not be particularly important for us. The reduced density matrix is then given by: 
\be
\rho^E_A = \text{Tr}_{\bar{A}}\;\rho^E = \oint d\beta\; e^{\beta E}\; Z_{\bar{A}}(\beta) \sum_i e^{-\beta E_i} |E_i \rangle \;\langle E_i|\
\ee
The trace giving the Renyi entropy is then equal to: 
\be 
\text{Tr}_A\;\left(\rho^E_A\right)^n = \left[\prod^n_{k=1} \oint d\beta_k\; e^{\beta_k E}\; Z_{\bar{A}}\left(\beta_k\right)\right]\times Z_A \left(\sum^n_{k=1}\beta_k\right)
\ee
In the saddle point approximation, the inverse Laplace transform is done by finding the saddle-points for  $\beta^*_k$. To be consistent with the cosmic-brane prescription,  we further assume that the saddle points are all identical:
\be
\beta^*_1 = \beta^*_2 = ... = \beta^*_n = \beta^*
\ee
In this case, it can derived that the saddle-point equation for the single parameter $\beta^*$ takes the form:
\be
\frac{\partial Z_{\bar{A}}(\beta)}{\partial \beta}\Big|_{\beta^*} + \frac{\partial Z_A(n\beta)}{n\partial \beta}\Big|_{\beta^*} = E  
\ee
Using the relation between subsystem and total energy in the high density limit: 
\be
\frac{\partial Z_{\bar{A}}(\beta)}{\partial \beta}\Big|_{\beta^*} = (1-f)\langle E\rangle_{\beta^*},\;\;\frac{\partial Z_A(n\beta)}{n\partial \beta}\Big|_{\beta^*}  = f\langle E\rangle_{n\beta^*}
\ee
We finally obtain the gluing/matching conditions (\ref{eq:canonical_match},\;\;\ref{eq:micro_match}) used in the gluing construction for computing the Renyi entropies in the micro-canonical ensembles. 

\section{Details of computing $n_{cut}$ }
\renewcommand{\theequation}{B.\arabic{equation}}
\setcounter{equation}{0}
\label{app:instabilities}
We first examine the onset of second order instabilities towards nearby one-zone black holes. As $n$ is varied, $\chi_1 \equiv \chi^{\bar{A}}_1$ is fixed, and we treat it as representing the temperature $T$. In the near-primary limit $n_c-1\sim \epsilon \ll 1$, it can be derived that the BTZ segment along A becomes unstable when: 
\be
\chi_1 n_c^{5}>4,\;\;\;n<\text{min}\left\lbrace\frac{\chi_1}{4} n_c^5,\;n_A\right\rbrace
\ee
The BTZ segment along $\bar{A}$ becomes perturbatively unstable when:
\be
\chi_1>4,\;\;\;n<n_{\bar{A}} 
\ee
The values $n_{A,\bar{A}}$ mark the thresholds where the BTZ segments become deformable in their respective GGEs, i.e. $\chi^{A,\bar{A}}_{2} = \tilde{\zeta}^{A,\bar{A}}_4$. They are explicitly given by:
\bea
n_A\left(\chi_1\right) &=& n_c^5 \chi_1 \left(\frac{\sqrt{6-4n_c^2-2n_c^4+4n_c^4 \chi_1}-2n_c^2+2}{1+2n_c^2-3n_c^4+2n_c^4\chi_1}\right)^2\nonumber\\
n_{\bar{A}}\left(\chi_1\right) &=& \frac{2n_c^5 \chi_1}{3-2n_c^2\left(1+2\sqrt{\chi_1}\right)+n_c^4\left(2\chi_1+4\sqrt{\chi_1}-1\right)}
\eea
In the high temperature regime, the two onset values $n_{A,\bar{A}}$ admit perturbative expansions in $\eta=1/\chi_1^{1/4}$: 
\bea\label{eq:high_T_p}
n_A\left(\chi_1\right) &=& n_c-\frac{2(n_c^2-1)}{n_c} \eta^2 + \frac{7n_c^4-10 n_c^2+3}{2n_c^3} \eta^4 +...\nonumber\\
n_{\bar{A}}\left(\chi_1\right)&=& n_c -\frac{2(n_c^2-1)}{n_c} \eta^2 + \frac{9 n_c^4-14 n_c^2+5}{2n_c^3} \eta^4+...
\eea
In this limit, they differ at the $\eta^4$ order and satisfy $n_{\bar{A}}>n_A$. On the other hand, the second order instabilities can only exist for sufficiently high temperature $\left(\chi_1 \geq 4 n_c^{-5}\right)$ and so is absent in the low temperature limit. 

Next we examine the first-order instabilities. The onset of such instabilities on a BTZ black hole with $\mathcal{Q}_1$ charge density $h$ is marked by the existence of another root $h'\neq h$ that simultaneously solves the following two equations: 
\be
\mathcal{G}(h')=\mathcal{G}(h),\;\;\mathcal{F}_{BTZ}(h')=\mathcal{F}_{BTZ}(h) 
\ee 
By eliminating $h'$, we obtain the following equation that controls the onset: 
\be\label{eq:BTZ_tunnel}
64 \chi_1^2+ \left(18-24 \chi_2-16 \chi_2^2\right)\chi_1+\left(2+\chi_2\right)\left(3+\chi_2\right)^3=0
\ee 
At the transition point (\ref{eq:BTZ_tunnel}), the other root $h'$ can be obtained from: 
\be 
{\frac{h'}{h}}=-\frac{7+4\chi_2+\sqrt{-15-8\chi_2}}{8}.
\ee 
Solutions of (\ref{eq:BTZ_tunnel}) are identified as the physical onset of the first order instabilities if $h'>0$. The onset $n'_A$ along the segment $A$ is the value of $n$ such that $(\chi^A_1,\;\chi^A_2)$ satisfy (\ref{eq:BTZ_tunnel}); while the onset $n'_{\bar{A}}$ along $\bar{A}$ is when $(\chi^{\bar{A}}_1,\;\chi^{\bar{A}}_2)$ satisfy (\ref{eq:BTZ_tunnel}). Explicit expressions for $n'_{A,\bar{A}}$ are expectedly very complicated and not particularly illuminating for generic temperature $T$ or $\chi_1$. Instead, we study their asymptotic behaviors in a few limits.  

In the high temperature limit $\chi_1 \gg 1$, the onset values $n'_{A,\bar{A}}$ also admit series expansions in $\eta$ analogous to (\ref{eq:high_T_p}): 
\bea\label{eq:high_T_f}
n'_A(\chi_1) &=& n_c - \frac{2\sqrt{2}(n_c^2-1)}{n_c} \eta^2 - \frac{2^{5/4}(n_c^2-1)}{n_c^2} \eta^3 + \frac{(n_c^2-1)(22n_c^2-21)}{4n_c^3} \eta^4+ ...\nonumber\\
n'_{\bar{A}}(\chi_1) &=& n_c - \frac{2\sqrt{2}(n_c^2-1)}{n_c} \eta^2 - \frac{2^{5/4}(n_c^2-1)}{n_c^2} \eta^3 - \frac{(n_c^2-1)(27n_c^2-26)}{4n_c^3} \eta^4+ ...
\eea
They differ at the $\eta^4$ order. Comparing (\ref{eq:high_T_p}) and (\ref{eq:high_T_f}), we conclude that the cut-off $n_{cut} = \text{max}\lbrace n_{A,\bar{A}},n'_{A,\bar{A}} \rbrace$ for the Renyi index is given by: 
\be
n_{cut}(\chi_1) =  n_{\bar{A}}(\chi_1) = n_c -\frac{2(n_c^2-1)}{n_c} \eta^2 + \frac{9 n_c^4-14 n_c^2+5}{2n_c^3} \eta^4+...
\ee

As the low temperature limit is approached, the value of $n'_A$ first becomes divergent towards the lower limit temperature $\chi_1 \to \Delta \chi_1$:
\be\label{eq:low_T_A}
n'_A (\chi_1)= \frac{17n_c^4-14 n_c^2+3-6n_c^6}{2n_c^3\left(\chi_1-\Delta \chi_1\right)}+...,\;\;\;\Delta\chi_1=\frac{(n_c^2-1)(3n_c^2-1)}{2n^4}
\ee 
Below this temperature, the BTZ segment along $A$ is unstable in the corresponding GGE for all Renyi index $n\geq 1$\footnote{The other onset value $n'_{\bar{A}}$ also exhibits a  divergence of similar nature at $\chi_1 = \Delta \bar{\chi}_1<\Delta \chi_1$, where the glued BTZ solution is already invalid for all $n\geq 1$.}. We conclude that for $\chi_1 \leq \Delta \chi_1$ the glued BTZ solution does not give the refined Renyi entropy $\tilde{S}_n(A)$ for all $n\geq 1$. Since for $\chi_{1}\sim \Delta \chi_1<4 n_c^{-5}$ the second order instability is absent, the cut-off Renyi index $n_{cut}$ near the lower limit $\Delta \chi_1$ is dictated by the first order instability: 
\be
n_{cut}(\chi_1) = n'_A(\chi_1)=\frac{17n_c^4-14 n_c^2+3-6n_c^6}{2n_c^3\left(\chi_1-\Delta \chi_1\right)}+...
\ee
  
There is an interesting intermediate regime for temperatures close to $\chi_1 \sim \chi_c$:
\be
\chi^{\bar{A}}_1 \approx \chi_c = \frac{1}{2}n_c^{-4}(1+n_c)^3
\ee
The temperature $\chi_c$ is marked by the property that $n'_A = n'_{\bar{A}} =1$ at this point. As a result the glued BTZ solution is free from such instabilities all the way down to $n=1$. Intuitively $\chi_c$ corresponds to the fine-tuned temperature such that the two local minimal $(h,\bar{h})$ of $\mathcal{F}_{BTZ}$ in the GGE $(\beta,\mu^3_1,\mu^1_1)$ have equal free energies. In the vicinity of $\chi_c$, the onset values $n'_{A,\bar{A}}$ are given by:
\bea
n'_A(\chi_1) &=& 1- \frac{2n_c^3(n_c-1)^2}{(n_c+1)^3(3-n_c)}\left(\chi_1-\chi_c\right)+...\nonumber\\
n'_{\bar{A}}(\chi_1) &=& 1+ \frac{2n_c^3(n_c-1)^2}{(n_c+1)^4}\left(\chi_1-\chi_c\right)+...
\eea
There remains the second order instabilities against nearby one-zone black holes. Those from the segment along $A$ are present only for $\chi_1>4 n_c^{-5}$; while those from the segment $\bar{A}$ are present for $\chi_1>4$. It is observed that:
\be
4 n_c^{-5}< \chi_c < 4
\ee 
As a result, in the vicinity of $\chi_c$ only the segment along $A$ is susceptible to second order instabilities. Therefore its onset value $n_A$ dictates the cut-off $n_{cut}$ at $\chi_c$, and can be expanded in small $n_c -1 \sim \epsilon$: 
\be
n_{cut}(\chi_c)= n_A(\chi_c) =  1+ \frac{5}{16}\left(n_c-1\right)^3+...
\ee
We see that $n_{cut}(\chi_1)-1$ is suppressed to the $(n_c-1)^3$ order near $\chi_1= \chi_c$, this is to be compared with the $(n_c-1)$ order in the asymptotically high temperature limit $\chi_1 \to \infty$. It shows that in the vicinity of the intermediate temperature $\chi_c$, the range of validity for the glued BTZ solution, and thus the result \eqref{eq:KdV_renyi}, extends the closest to $n=1$. 

\section{Toy model analysis for entanglement spectral density }
\renewcommand{\theequation}{B.\arabic{equation}}
\setcounter{equation}{0}
\label{app:ent_spectrum}
To this end, we can work with the following expression for $S_n(A)$ as a toy model, whose full $n$-dependence effectively captures the qualitative behaviors just listed: 
\bea
S_n(A) &=&  f S_{th} + \left(\frac{\delta n_{cut}\Delta S}{1-n}\right) \ln{\left[\cosh{\left(\frac{n-1}{\delta n_{cut}}\right)}\right]} 
\eea
We need to evaluate the inverse Laplace transform: 
\be
g(\lambda) = \left(\frac{\delta n_{cut}\; e^{\lambda}}{2\pi i}\right)\times\int^{\Gamma+i\infty}_{\Gamma-i\infty} dx \;e^{x\; \delta n_{cut}(\lambda-fS_{th})}\cosh{\left(x\right)}^{\delta n_{cut}\Delta S} ,\;\;\;x = \left(\frac{n-1}{\delta n_{cut}}\right) 
\ee
This is still difficult for general choices of parameters. We can make further progress by assuming that 
\be 
M=\delta n_{cut}\Delta S \sim \sqrt{q_1} \in \mathds{N}
\ee
is a very large integer. In this case one can use the binomial expansion to obtain a series of delta-function peaks: 
\be
g(\lambda) = \frac{e^{\lambda}}{2^{M}} \sum^{\Delta S}_{m=0}\left( \begin{array}{c}
M \\
m\\
\end{array}\right) \delta \left(\lambda- \lambda_m\right),\;\;\lambda_m = fS_{th}-\delta n_{cut}^{-1}\left(M
-2m\right)  
\ee
The spectrum $\lbrace \lambda_m\rbrace$ is confined to the interval: 
\be\label{eq:spectrum_range} 
\lambda_m \in \left[fS_{th} -\Delta S, \;f S_{th} +\Delta S\right]
\ee
In the holographic and high density limit $S_{th} \sim \Delta S \sim \pi\sqrt{q_1}/2G_N\to \infty$, we can bin together the $\delta$-function peaks into a continuous binomial distribution inside (\ref{eq:spectrum_range}) weighted by an exponential growth factor. This produces the entanglement spectral density $g(\lambda)$ plotted in Figure (\ref{fig:renyi_to_spectrum}). 

\bibliographystyle{JHEP} 
\bibliography{ref}

\end{document}